\documentclass[sigconf,screen]{acmart}

\makeatletter
\def\subsubsection{\@startsection{subsubsection}{3}%
  \z@{.5\linespacing\@plus.7\linespacing}{.1\linespacing}%
  {\it\sf\bf}}
\makeatother

\makeatletter
\renewcommand\@formatdoi[1]{\ignorespaces}
\makeatother
\renewcommand\footnotetextcopyrightpermission[1]{}

 \renewcommand\footnotetextcopyrightpermission[1]{} % removes footnote with conference information in first column
\usepackage{multirow}
\usepackage{graphicx}
\usepackage[linesnumbered,ruled,vlined]{algorithm2e}
\usepackage{epstopdf}
\definecolor{codegreen}{rgb}{0,0.6,0}
\definecolor{codegray}{rgb}{0.5,0.5,0.5}
\definecolor{codepurple}{rgb}{0.58,0,0.82}
\definecolor{backcolour}{rgb}{0.95,0.95,0.92}
\definecolor{myxcodecolour}{rgb}{0.682,0.059,0.588}
\usepackage{listings}
\lstdefinelanguage{GSQL}{
    keywords={CREATE, QUERY, FROM, SELECT, WHERE, ACCUM, PRINT, SumAccum, VERTEX, ALTER, EMBEDDING, ATTRIBUTE, SPACE, PRIMARY, KEY, LOAD, VALUES, VectorSearch,ORDER, BY, LIMIT, ADD },
    keywordstyle=\color{codepurple}\bfseries,
    morecomment=[l]{--},
    morecomment=[l]{//},
    morecomment=[s]{/*}{*/},
    commentstyle=\color{codegreen}\itshape,
    morestring=[b]",
    sensitive=true,
    stringstyle=\color{red},
}

\newcommand{\myline}[1]{{\medskip\noindent\textbf{#1.}}}

\usepackage{url}
\usepackage[labelfont=bf]{caption}
\usepackage{xcolor}
\usepackage{color, colortbl}

\usepackage{pifont}

\newcommand{\sys}{TigerVector}
\newcommand{\sect}{Sec.}
\newcommand{\shige}[1]{{\it\small\textcolor{blue}{[ {#1}\ --shige ]}}}
\newcommand{\mingxi}[1]{{\it\small\textcolor{red}{[ {#1}\ --mingxi ]}}}
\newcommand{\yuxu}[1]{{\it\small\textcolor{orange}{[ {#1}\ --yu]}}}

\begin{document}
\pagestyle{empty} % removes running headers

% \title{TigerVector: Supporting Vector Data Analytics in Graph Databases}
% \title{TigerVector: Supporting Vector Data Analytics in Graph Databases with "Edge-as-Index"}
%\title{Supporting Large-scale Vector Data Analytics in Graph Databases with "Edge-as-Index"}
% \title{Supporting Large-scale Vector Data Analytics in Graph Databases}
% \title{When Graph-based Vector Search Meets Graph Databases}

% \title{Graph+Vector RAG: Integrated Vector Search in Graph Databases for Advanced RAG}

\title{TigerVector: Supporting Vector Search in Graph Databases for Advanced RAGs}

 \author{Shige Liu}
% % \orcid{0000-0000-0000-0000}
 \affiliation{%
   \institution{Purdue University}
 }
 \email{liu3529@purdue.edu}

 \author{Zhifang Zeng}
% % \orcid{0000-0000-0000-0000}
 \affiliation{%
   \institution{TigerGraph}
 }
 \email{michael.zeng@tigergraph.com}

 \author{Li Chen}
% % \orcid{0000-0000-0000-0000}
 \affiliation{%
   \institution{TigerGraph}
 }
 \email{li.chen@tigergraph.com}

 \author{Adil Ainihaer}
% % \orcid{0000-0000-0000-0000}
 \affiliation{%
   \institution{TigerGraph}
 }
 \email{adil.ainihaer@tigergraph.com}

 \author{Arun Ramasami}
% % \orcid{0000-0000-0000-0000}
 \affiliation{%
   \institution{TigerGraph}
 }
 \email{arun.ramasami@tigergraph.com}

 \author{Songting Chen}
% % \orcid{0000-0000-0000-0000}
 \affiliation{%
   \institution{TigerGraph}
 }
 \email{songting.chen@tigergraph.com}

 \author{Yu Xu}
% % \orcid{0000-0000-0000-0000}
 \affiliation{%
   \institution{TigerGraph}
 }
 \email{yu@tigergraph.com}

 \author{Mingxi Wu}
% % \orcid{0000-0000-0000-0000}
 \affiliation{%
   \institution{TigerGraph}
 }
 \email{mingxi.wu@tigergraph.com}

 \author{Jianguo Wang}
% % \orcid{0000-0000-0000-0000}
 \affiliation{%
   \institution{Purdue University}
 }
 \email{csjgwang@purdue.edu}

\begin{abstract}
%\mingxi{short version}
In this paper, we introduce \sys{}, a system that integrates vector search and graph query within TigerGraph, a Massively Parallel Processing (MPP) native graph database. We extend the vertex attribute type with the \texttt{embedding} type. To support fast 
vector search, we devise an MPP index framework that interoperates efficiently with the graph engine. The graph query language GSQL is enhanced to support vector type expressions and enable query compositions between 
vector search results and graph query blocks. These advancements elevate the expressive power and analytical capabilities of graph databases, enabling seamless fusion of unstructured and structured data in ways previously unattainable. Through extensive experiments, we demonstrate \sys{}'s hybrid search capability, scalability, and superior performance compared to other graph databases (including Neo4j and Amazon Neptune) and a highly optimized 
% \yuxu{and} % Jianguo: should be fine : ) 
specialized vector database (Milvus). \sys{} was integrated into \textbf{TigerGraph v4.2}, the latest release of TigerGraph, in December 2024.
%  \jianguo{Will proofread}
\end{abstract}

\maketitle

\footnotetext{ This paper is accepted to appear in the SIGMOD 2025 Industrial Track. The final version will be published by ACM.}

\section{Introduction}\label{sec:intro}

Retrieval-augmented generation (RAG) is an emerging technology that grounds LLMs to enhance accuracy and reliability by retrieving relevant context from external data sources~\cite{FanDNWLYCL24}. Today, most RAGs are based on vector databases~\cite{Milvus21,SingleStoreV}, which store the semantic embeddings of  underlying data. However, vector-based RAGs fall short in many complex applications, as they cannot capture the accurate relationships between underlying data objects. 
% As a result, 
Thus, vector-based RAGs frequently suffer from poor prompt hit rates~\cite{arizeRAG,Gartner5188263}, particularly in scenarios where the retrieved context does not fully align with user queries. This can result in multiple iterative LLM API calls to refine the response, increasing both latency and costs. 

% These inefficiencies underscore the need for more advanced approaches that integrate complementary retrieval methods. % This sentence is not well matched with the next paragraph
% \mingxi{add "As a result.." the economic(\$\$) challenge now with vector db}

To address these challenges, a new concept called GraphRAG was recently proposed~\cite{edge2024local,abs-2408-08921,Neo4jRAG2}. Instead of using vector databases, GraphRAG leverages graph databases (e.g., Neo4j~\cite{Neo4jRAG2} or TigerGraph~\cite{tigergraph}) to store knowledge graphs of underlying data. This approach is useful for answering questions that may require understanding information across multiple documents, such as finding all positive reviews written by a specific customer or summarizing the impact of COVID-19 on the global economy.

In this paper, we advocate for a hybrid RAG, \textit{VectorGraphRAG}, which combines vector-based and graph-based RAGs to leverage the best of both worlds. This approach will enable many new possibilities for grounding LLMs by leveraging both vector search and graph search~\cite{Neo4jRAG2}. For example, one could perform vector search and graph search separately to retrieve different candidate sets and merge them to obtain a comprehensive result set. One could also use vector search to identify a smaller set of results first and then apply graph traversal to expand it for more relevant context.
% for LLMs.

% \begin{sloppypar}
In particular, we aim to support VectorGraphRAG by developing \sys{},  
% \textcolor{blue}{as data retrieval infrastructure}, 
which integrates vector search seamlessly into TigerGraph, a distributed graph database system.

% \mingxi{this paragraph is needed? Can we remove "In this paper"? it's the same as the previous paragraph "In this paper"}\jianguo{Yes, the previous paragraph introduces VectorGraphRAG. But this paragraph says what we do for VectorGraphRAG: integrating vector search inside graph DB. It's a quick overview of what we do (as some reviewers may skip abstract).}
% \end{sloppypar}

While the straightforward way to support VectorGraphRAG is to use two separate databases (i.e., a vector database and a graph database), we argue that it is highly desirable to have a \textit{unified system} that natively supports both vector data and graph data. First, it reduces data movement and minimizes data silos by keeping both vector data and graph data within the same system. Second, it maintains data consistency by directly linking the vector embeddings with their source data. Third, it facilitates hybrid searches of vector and graph data using a unified query language, which is GSQL~\cite{GSQL} in our case. Fourth, it supports efficient data governance by providing a single set of access controls (e.g., role-based access control) for both vector data and graph data. %\yuxu{replace RBAC with Role Based Access Control, since it appears only one time}

\myline{Challenges} It is challenging to efficiently support vector search within graph database systems. (1) Since vector data is not a first-class citizen in graph databases, it is non-trivial to achieve high query performance compared to specialized vector database systems like Milvus~\cite{Milvus21}. (2) Since graph databases utilize a graph query language, it remains unclear how to represent different types of vector search queries (such as pure vector search, filtered vector search, and hybrid vector and graph search) and execute them efficiently. (3) It is crucial to support updates that are both transactionally consistent and efficient, ensuring that updates to vector data do not negatively impact other graph data. (4) It is challenging to efficiently manage different types of vector embeddings (e.g., text embeddings or image embeddings) within graph databases.
% \end{enumerate}

\myline{Limitations of Prior Work} Although a few graph databases (e.g., Neo4j~\cite{Neo4jVec} and Amazon Neptune~\cite{NeptuneVec}) have started to support vector search, they address the above challenges only partially and have many limitations. For example, Neo4j and Neptune lack high performance, fail to support advanced vector search (e.g., filtered vector search and hybrid search), do not provide efficient and atomic updates, and lack support for multiple embedding types.

\myline{Overview of \sys{}} In this paper, we present \sys{}, a new system that adds efficient vector search support within TigerGraph, a distributed graph database system. It introduces vector embeddings as a new attribute of existing graph nodes, similar to other attributes within the graph nodes. Moreover, it supports multiple types of vector embeddings within the same graph node. The vector index is based on HNSW~\cite{HNSW}, 
which is widely used in modern vector databases. \sys{} introduces a suite of techniques to address the above challenges as described below.

To address the performance challenge, \sys{} fully leverages TigerGraph’s MPP architecture, allowing multiple compute cores or machines to process queries in parallel. Moreover, \sys{} decouples the storage of vector embeddings from other graph attributes to best utilize the native vector index implementation.

To address the usability challenge, \sys{} integrates vector search into GSQL~\cite{GSQL},  
TigerGraph’s SQL-like declarative graph query language. 
In particular, it supports both pure vector search and advanced vector search, including filtered vector search, vector search on graph patterns, and vector similarity join on graph patterns. Moreover, \sys{} supports flexible vector search functions that seamlessly interact with other GSQL features such as vertex set variables and graph algorithms.
%\yuxu{ "for customized queries" doesn't read right (since one can say every customer query is  "customized" with or without vector search. Need to rework on this sentence}.\jianguo{Sure, shige can change it}

To address the update challenge, \sys{} supports both efficient and atomic updates. Specifically, the decoupling of vector embeddings from other attributes in graph nodes allows \sys{} to handle incremental vector index updates effectively. This design also facilitates graph updates and reduces update overhead. 
Moreover, updates involving both graph attributes and vector attributes are performed atomically.

For efficient vector embedding management, \sys{} introduces a new data type called \texttt{embedding} to handle vector embeddings decoupled from other graph attributes. More importantly, the \texttt{embedding} type not only manages important metadata, such as dimensionality and distance metric, but also enables the definition of multiple embedding types.

% \shige{\sys{} uses \texttt{embedding} type to manage important metadata, such as dimensionality and distance metric, and provides a convenient way to define multiple embedding types.}

% \vspace{-0.1cm}
\myline{Experimental Overview}
%Experiments on various datasets, including SIFT100M and Cohere10M, show that we achieved comparable performance with Milvus, the state-of-the-art specialized vector database, and outperforms the vector search performance of Neo4j, the leading \yuxu{change leading to well known} graph database with embedding support, with 1.65X - 2.6X QPS and 22\% - 26\% higher recall rate. \mingxi{talking about QPS without the same recall is not a fair comparison. I would suggest articulate Neo4j cannot allow you adjust recall and at 60\% recall, TG outperform Neo4j. Plus TG can scale out, and achieve on par or better performance than Milvus on billion scale vectors}
Experiments on SIFT100M and Deep100M show that \sys{} significantly outperforms both Neo4j and Amazon Neptune in vector search. Specifically, \sys{} achieves \textbf{3.77}$\times$ to \textbf{5.19}$\times$ higher throughput and \textbf{23\%} to \textbf{26\%} higher recall rate compared to Neo4j. Moreover, \sys{} achieves \textbf{1.93}$\times$ to \textbf{2.7}$\times$ higher throughput at significantly lower hardware costs (\textbf{22.42}$\times$ less) compared to Amazon Neptune while maintaining a similar recall rate. Besides that, \sys{} also achieves comparable, and sometimes even higher, throughput compared to Milvus.

% \vspace{-0.2cm}
\myline{Contributions} The main contribution of this work is \sys{}, a novel system that efficiently supports vector search within TigerGraph, a distributed graph database system. 
It introduces a suite of new techniques (see \sect{}~\ref{sec:design}) to address the challenges in performance and usability. Extensive experiments show that \sys{} can achieve comparable and even higher performance than Milvus, a highly optimized 
% \yuxu{and} 
specialized vector database. \sys{} also significantly outperforms other graph databases, 
such as Neo4j and Amazon Neptune, in vector search performance. Moreover, \sys{} supports advanced vector search efficiently. \sys{} was integrated into \textbf{TigerGraph v4.2} in December 2024. \textcolor{black}{The community edition can be downloaded via \url{https://dl.tigergraph.com/}}.

Although \sys{} is designed based on TigerGraph, many of the proposed design decisions and techniques are applicable to other graph databases. Further discussion can be found in \sect{}~\ref{sec:discussion}.

% \myline{Paper Organization} \sect{}~\ref{sec:background} provides an overview of the background and related work. \sect{}~\ref{sec:design} outlines the overall system design. \sect{}~\ref{sec:indexdesign} details the vector index design. \sect{}~\ref{sec:query_processing} describes the vector search design. \sect{}~\ref{sec:experiment} presents the experimental results. \sect{}~\ref{sec:discussion} discusses key lessons learned and the applicability to other graph databases. \sect{}~\ref{sec:conclusion} concludes the work.

% \vspace{-1cm}

\section{Background and Related Work}\label{sec:background}

% In this section, we provide an overview of TigerGraph and related work.

\subsection{TigerGraph}\label{sec:tigergraph_background}

Graph databases have emerged as 
% the third canonical 
a new database category.
% alongside relational databases and key-value databases. 
Prominent commercial graph databases include TigerGraph~\cite{tigergraph}, Neo4j~\cite{Neo4j}, and Amazon Neptune~\cite{Neptune}. 
% \yuxu{change "the third" to a new, and add document databases to second part} % Jianguo: done
% 
% A native graph database is distinguished by its storage format, which organizes data as nodes (vertices) and edges (relationships). \yuxu{the above sentence isn't clear or sufficient. People may argue RDBMS can organize graph data as node tables and edge tables. }  To optimize storage efficiency, compression techniques are typically employed. \yuxu{all databases use compressions, so the above sentence alone is moot. The last two sentences don't read well. try something like below.}
% \yuxu{A native graph database is purpose-built for managing graph data, with nodes, edges, and properties directly represented in its storage architecture. Relationships (edges) are treated as first-class entities, typically implemented as direct pointers or links between nodes, enabling highly efficient traversal operations. Beyond traditional data compression techniques, native graph databases also employ graph-specific compression methods optimized for their structures.} % Jianguo: done.
%
A native graph database is purpose-built for managing graph data, with nodes, edges, and properties directly represented in its storage architecture. Relationships (edges) are treated as first-class entities, typically implemented as direct pointers or links between nodes, enabling highly efficient traversal operations. Beyond traditional data compression techniques, native graph databases also employ graph-specific compression methods optimized for their structures.

Graph databases excel in handling graph-shaped query workloads, offering significant performance advantages over other database types. Benchmark suites like LDBC-SNB, LDBC-FinBench, and LDBC-GraphAnalytics measure performance for such workloads~\cite{LDBC}.
The key performance advantage of graph databases lies in their design: common joins are materialized as edges during data ingestion. At runtime, the query engine traverses these edges to access related data directly, often eliminating the need for expensive join operations.

TigerGraph employs the property graph model~\cite{AnglesBD0GHLLMM23,Tian22}, representing data through vertex and edge types. Its graph model allows both directed and undirected edges, and allows multiple edges (of the same type or different types) between two nodes. 
% Its graph model supports mixed edge types (both directed and undirected), allows multiple edges  between two nodes, and requires each vertex or edge to adhere to a predefined schema.   \yuxu{ try use the following to replace the above sentence. }   \yuxu{Its graph model allows both directed and undirected edges, and allows multiple edges (of the same type or different types) between two nodes.}  % Jianguo: done  

TigerGraph is a distributed, native graph database designed for scalability and high performance. It supports deployment on clusters, with data automatically partitioned and distributed across nodes. 
In the storage layer, data is organized into units called \textit{segments}, each containing a fixed number of vertices. These segments serve as the fundamental units for parallel and distributed computing. 
% parallel computation  and are distributed across machines. 
Outgoing edges are stored within the source vertex’s segment.
% In the storage layer, data is organized into parallel units called \textit{segments}, each with the same vertex count. Outgoing edges are stored within the source vertex’s segment. 
% \yuxu{try something like below to replace the above sentence}
% \yuxu{In the storage layer, data is organized into units called segments, each containing a fixed number of vertices. These segments serve as the fundamental units for parallel computation.}. Outgoing edges are stored within the source vertex’s segment. } % Jianguo: done
The compute layer leverages a massively parallel processing 
% \yuxu{remove:(MPP) }
architecture to enable hybrid transactional and analytical processing. Two key parallel primitives, \textit{VertexAction} and \textit{EdgeAction}, allow user-defined functions to operate across segments in parallel. 
% \yuxu{jianguo: are bold fonts typically used here for EdgeAction/VertexAction. It just looks a little abrupt here.} % Jianguo: Changed to non-bold.

\begin{comment}
\begin{itemize}
    \item Mixed edge types: Both directed and undirected edges are allowed.
    \item Multigraph support: Multiple edges can exist between two nodes.
    \item Closed schema: All nodes and edges must adhere to a predefined schema.
\end{itemize}
\end{comment}

TigerGraph’s query language, GSQL~\cite{GSQL}, is a Turing-complete graph database query language. One of its key advantages over other graph query languages is its support for accumulators~\cite{GSQL}, which can be either global or vertex-local. It also includes flow control primitives such as \texttt{WHILE}, \texttt{IF-ELSE}, \texttt{FOREACH}, and others to support graph algorithms. GSQL enables query block composition using vertex set variables and accumulators. A query is composed of a sequence of query blocks (\texttt{SELECT-FROM-WHERE}), each producing a vertex set variable. 
% \yuxu{remove:In top-down syntax order, subsequent } % Jianguo: ok 
Query blocks in the \texttt{FROM} clause can utilize vertex set variables from prior blocks, thus facilitating query block composition. Additionally, GSQL supports \texttt{UNION}, \texttt{INTERSECT}, and \texttt{MINUS} binary operators between two vertex set variables.

Accumulators can be used as another compositional tool in GSQL.   %
% \shige{mentioned accumulator in previous paragraph}. 
They are runtime variables that are mutable throughout the query life cycle. Global accumulators (prefixed by \texttt{@@}) can be read and written within a query block or across different query blocks. Vertex-local accumulators (prefixed by \texttt{@}) can be read and written within a query block when their associated vertex is accessible. Since vertices can be activated in different query blocks, query composition is facilitated through their attached runtime vertex accumulators. 
% \yuxu{overall I feel this paragraph doesn't read well and we'd lose points here. It doesn't strengthen the paper and readers may feel it's a stretch to talk about  accumulator and composition-ability. We can either remove this paragraph or just use a few sentences from \cite{GSQL} to describe/define global and local Accum.  }% Jianguo: we can keep it for now
% \cite{GSQL}  

% \yuxu{less is better. I suggest we just copy  the def of local and global ACCUM from the GSQL paper. Wechated the screenshot to you guys.   The current paragraph doesn't read well 
% }
%\shige{I think current is good. }
\subsection{Vector Databases}

Vector databases~\cite{Milvus21,SingleStoreV,VecDBSurvey24,VecDBTutorial24,VecDBPanel24} have recently gained significant attention due to their important 
% \yuxu{remove?  critical } % Jianguo: should be fine
role in grounding LLMs through vector-based RAGs.

There are two types of vector databases in the literature: specialized vector databases and integrated vector databases. Specialized vector databases are designed explicitly to manage vector data, achieving high performance and scalability. Examples include Milvus~\cite{Milvus21}, Pinecone~\cite{Pinecone}, and Weaviate~\cite{Weaviate}. Integrated vector databases aim to incorporate vector search into existing data systems, such as relational databases~\cite{VecDBRDBMS} and graph databases~\cite{Neo4jVec}. Examples include pgvector~\cite{Pgvector}, PASE~\cite{PASE20}, and SingleStore-V~\cite{SingleStoreV}.

This work follows the integrated approach because real-world applications often involve more than just vector data. A unified system can address various issues, such as data consistency, data silos, and advanced vector search, as discussed in \sect{}~\ref{sec:intro}.

\subsection{Supporting Vector Search in Graph Databases}

To the best of our knowledge, only a few graph databases, including Neo4j~\cite{Neo4jVec}, Amazon Neptune~\cite{NeptuneVec}, and DGraph~\cite{DgraphVector}, support vector search. However, their support is rudimentary and comes with many limitations. For example, Neo4j implements vector search using Lucene's index~\cite{Neo4jVecLuceneIndex}. However, it does not 
% \shige{fully} 
support index parameter tuning, which is crucial in vector databases to achieve high performance. It also lacks support for advanced vector search, declarative vector search using a graph query language, and the ability to accommodate multiple embedding types. Our system addresses all of these limitations.

% \yuxu{remove:Moreover,} % Jianguo: done
Amazon Neptune~\cite{NeptuneVec} and DGraph~\cite{DgraphVector} share the limitations mentioned above. In addition, Neptune explicitly states that updates to the vector index are not atomic~\cite{NeptuneVec}, whereas \sys{} supports transactional updates.
% \yuxu{remove:with atomicity}. % Jianguo: done
Moreover, Neptune builds a single vector index for the entire graph, and this vector index is not distributed, which significantly limits its scalability for vector search~\cite{NeptuneVec}. In contrast, \sys{}'s vector index is distributed.
% \jianguo{This section can be expanded up to the end of this column}

% \newpage

\section{System Design Overview}\label{sec:design}

\begin{figure}[tbp]
\centering
\includegraphics[width=0.48\textwidth]{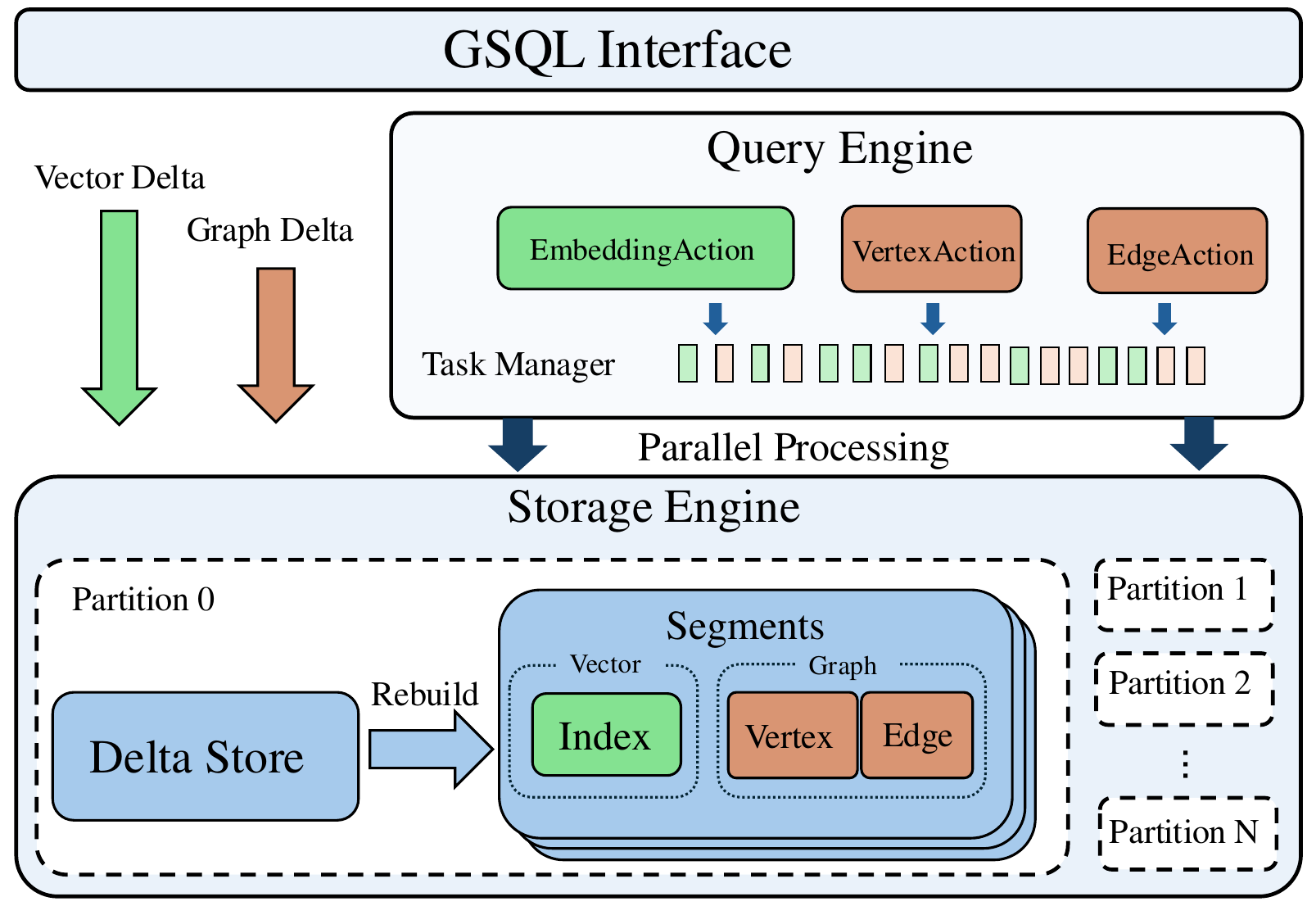}
\caption{System Overview}\label{fig:TigerVector_Overview}
% \vspace{-0.3cm}
\end{figure}

In this section, we present the system architecture of \sys{}, as illustrated in Figure~\ref{fig:TigerVector_Overview}. Our system is built inside TigerGraph, a distributed graph database, which partitions vertices into segments and evenly distributes these segments across multiple machines to fully utilize all compute resources. 
% \yuxu{remove (since it's already mentioned in the TG section):, a distributed graph database, which partitions vertices into segments and evenly distributes these segments across multiple machines to fully utilize all compute resources}. % Jianguo: we may keep it here, because people may directly read this section without reading TG section
\sys{} manages embeddings using a vertex-centric approach, storing all embeddings related to the same vertex segment together and building a vector index per segment, thereby unifying the management of vector data and graph data.

Since TigerGraph is a property graph database, where each node contains multiple key-value pairs as attribute properties, \sys{} introduces vector embeddings as a new attribute for existing graph nodes, similar to other attributes. Moreover, \sys{} supports multiple embedding types within the same graph node, as some applications may require different vectors for various types of data (e.g., document data or image data).

Next, we highlight the key features of \sys{}.

\myline{Embedding Type} Vectors are (float) data arrays generated by machine learning models that can capture the semantic meaning of the underlying data. Instead of managing vectors using data type \texttt{LIST<float>} as if they were simple lists of floats, 
% \yuxu{either remove the above "of data" or change "of data" to "of floats"?} %Jianguo: done
we introduce a new \texttt{embedding} data type to efficiently manage vectors and associated metadata, such as dimension, data type, and model information. We also introduce the concept of \texttt{embedding space}, which defines multiple \texttt{embedding} types from a specific machine learning model. More details can be found in \sect{}~\ref{sec:embedding_schema}.

\myline{Decoupled Storage for Vectors} Vector embeddings associated with the same vertex segments are stored as a single embedding segment and are stored separately from other attributes, allowing them to be managed independently of their vertex segments. 
Vector indexes are built for each embedding segment and are incrementally updated as new vector deltas are incorporated into embedding segments. Background vacuum processes are deployed for efficient updates while ensuring transaction support. The number of index update threads is dynamically tuned to strike a balance between efficiency and responsiveness for other queries.

%Efficient updates to vector indexes are supported while ensuring transactional consistency\shige{not transactional consistency}. To handle the mismatch between the graph and vector delta rebuild rates\jianguo{Need a revisit}\shige{not this logic. should be }, we allocate an appropriate number of threads to update vector indexes, ensuring efficient index building while maintaining responsiveness for other queries. More details can be found in \sect{}~\ref{sec:embedding_storage} and \sect{}~\ref{sec:incr-update}.

\myline{MPP Design} 
% \yuxu{MPP Design is good.  MPP-native is unnecessary and a new term} %Jianguo: ok, changed to MPP design
We fully leverage the MPP (Massively Parallel Processing) architecture of TigerGraph to achieve high performance. In this architecture, graphs and vectors are partitioned into vertex segments and embedding segments. Queries on large datasets are distributed as multiple tasks across these vertex and embedding segments, enabling parallel processing and distributed processing. 
% Tasks are further divided into sequential actions, including vertex/edge actions and vector actions, to mitigate the straggler effect. \yuxu{the last two sentences will  confuse the readers, I only did minimal editing. Maybe better just combine and reduce the last two senescence to a single one.}\jianguo{Agree, Shige can edit it first.} Jianguo: it's easier to delete it.
\sect{}~\ref{sec:vector_search} presents more details.
% Bitmaps are transferred between actions to indicate the vertices/vectors to process. The detailed query execution under this MPP architecture is explained in \sect{}~\ref{sec:vector_search}.

\myline{GSQL-integrated Declarative Vector Search} 
We integrate vector search into GSQL~\cite{GSQL} to support declarative vector search by extending the \texttt{ORDER BY...LIMIT} syntax.
More advanced queries, such as filtered vector
search, vector search on graph patterns, and vector similarity join on graph patterns, are also supported.
Detailed descriptions are provided in \sect{}~\ref{sec:query_processing}.

\myline{Flexible Vector Search Function} 
Our system also supports a powerful vector search function -- \texttt{VectorSearch()}, enabling flexible vector searches across multiple vertex types. It seamlessly integrates into GSQL queries by accepting a vertex set variable from prior query blocks for filtered searches and returning a vertex set variable for further composition. Optional parameters include an accumulator that returns top-k distances and corresponding vertices, and an index search parameter that adjusts search accuracy. By using vertex set variables as input and output, the function supports easy integration with graph algorithms, unlocking new avenues for advanced vector search applications.

\begin{comment}
Since this function will return a top-k vertex set, it can be seamlessly integrated into the query block composition flow that GSQL already has using vertex set variables. This function also takes an accumulator as an optional parameter, which can communicate the top-k distance and their corresponding vertex back to the query body for query composition.

Vertex set variables and accumulators can also be composed with the function for flexible queries. Accuracy is also adjustable using index search parameters. More importantly, our system can integrate graph algorithms with vector search functions to support a broader range of applications. More details are discussed in \sect{}~\ref{sec:query_processing}.
\end{comment}

% \yuxu{we can remove the following paragraph discussing in the cloud.  It doesn't add values to the paper, distracts the flow.} \jianguo{agree}
% \myline{Operating in the Cloud}\shige{Have to check with Li. May add more content.} We also support deployment on the cloud, where users can store data in S3 and start a variable number of machines for analytic tasks. Vector indexes, as well as other graph segment data, will be distributed to machines in a round-robin manner during the startup phase. A designated read/write worker handles data manipulation queries, while all other workers are read-only. Additionally, we minimize the amount of embedding rewrites on S3 to reduce the impact of S3 disk write operations.\jianguo{This paragraph is unclear.}

\section{Vector Index Design}\label{sec:indexdesign}

\subsection{Embedding Type}\label{sec:embedding_schema}

In \sys{}, we introduce a new data type called \texttt{embedding} to manage the vector attributes within graph vertices. It specifies metadata for the vectors, including dimensionality, the machine learning model used to generate the embedding, the vector index, the vector data type, and the similarity metric. Furthermore, each graph vertex can have one or more \texttt{embedding} attributes alongside other attributes, which provides finer granularity and more flexibility.

% \textcolor{blue}{which provides finer granularity and more flexibility compared to associating an embedding type with a graph in Amazon Neptune~\cite{NeptuneVec}.}

% \jianguo{Explain what's embedding schema.}
% In \sys{}, we introduce \shige{\texttt{embedding} type} as the description of a vector and its metadata information, including the dimensionality, the machine learning model used to generate the embeddings, the vector index, the vector data type, and the similarity metric.
% \shige{Delete: In \sys{}, we introduce a new data type called \texttt{embedding} to define the \textbf{embedding type} of vector attributes within vertices.} Also, each vertex can have one or more \texttt{embedding} attributes alongside other attributes.
% \jianguo{Need to specify where're the actual vectors stored. It seems the embedding type is only about the metadata not the actual vector data.}

Consider an example from the LDBC SNB (LDBC Social Network Benchmark)~\cite{ldbc-snb}, a standard benchmark for graph databases that models a social network application. It includes entities such as people, posts, comments, 
% \yuxu{posts, comments} 
and places, as well as their relationships and activities, including creates, replies, and likes. We can augment each \texttt{Post} node with a vector embedding representing its content, as follows.

%\mingxi{use Post( id INT PRIMARY KEY, author STRING...}
\begin{comment}
\begin{minted}[escapeinside=||]{SQL}
-- embedding type example
CREATE VERTEX Post (
id INоT PRIMARY KEY, author STRING, content STRING);

ALTER VERTEX Post 
ADD |\textcolor{myxcodecolour}{EMBEDDING}| |\textcolor{myxcodecolour}{ATTRIBUTE}| content_emb (
  DIMENSION = 1024,
  MODEL = GPT4,
  INоDEX = HNSW,
  DATATYPE = FLOAоT,
  METRIC = COSINE
);
\end{minted}
\end{comment}
\begin{lstlisting}[basicstyle=\fontsize{7.3pt}{8.4pt}\ttfamily]
-- embedding type example
CREATE VERTEX Post (
    id INT PRIMARY KEY, 
    author STRING, 
    content STRING
);

ALTER VERTEX Post 
ADD EMBEDDING ATTRIBUTE content_emb (
    DIMENSION = 1024,
    MODEL = GPT4,
    INDEX = HNSW,
    DATATYPE = FLOAT,
    METRIC = COSINE
);
\end{lstlisting}
% \jianguo{Why it's "Embedding Attribute" (nost "Embedding") in the above example? }

Defining the \texttt{embedding} type has many advantages compared to relying solely on existing data types such as \texttt{LIST<FLOAT>}. First, vectors are not merely simple arrays of floats. The metadata associated with vectors, such as dimensions, data types, and models, is also important and should be explicitly managed. For example, vector search across multiple vertex types is enabled using the \texttt{embedding} type, which allows searching multiple vector attributes generated by the same model. The key challenge in this search is to ensure compatibility, as vectors with different dimensions or generated by different models must be handled differently. We address this challenge through static analysis of the embedding data involved during query compilation, leveraging the metadata from \texttt{embedding} types. If all aspects of the vector metadata, except for the index type, are identical, the query is allowed. Otherwise, the query is rejected and a semantic error is returned.

% the query is rejected due to a semantic error. \yuxu{the query is rejected and a semantic error is returned.}

%\shige{better put this part to related work?However, it is not trivial to support it. Most of relational databases support strong type, but they only manage embedding a list of floats, which leaves them challenge of dynamic compatibility check; Neo4j is schema less, and does not support compatibility check on multiple vertex types; Neptune can even only support one index per graph, let alone multiple embedding attributes management.}
% \yuxu{ remove this sentence: However, if we use \texttt{LIST<FLOAT>} to store embeddings, the metadata would incur additional overhead for schema management. }  \yuxu { the sentence before this has good justification already. This extra sentence will ask for unnecessary questions from readers like: overhead of managing metadata is needed anyway, it doesn't really matter where you store the metadata.}\jianguo{agree}

Second, it simplifies the data loading process because vectors and other graph attributes often originate from different data sources, with vectors typically generated by a machine learning model. For example, we can easily load data from two separate files (\texttt{f1} and \texttt{f2}) to populate vectors and graph attributes using the following script, where vector values are assumed to be separated by ":". Otherwise, this process would be much more complex.

\begin{comment}
\begin{minted}[escapeinside=||]{SQL}
-- data loading example
CREATE loading job j1 FOR graph g1{
  LOAD f1 TO VERTEX Post VALUES (id, author, content); 
  LOAD f2 TO |\textcolor{myxcodecolour}{EMBEDDING}| |\textcolor{myxcodecolour}{ATTRIBUTE}| content_emb 
    ON VERTEX Post VALUES (id, split(content_emb, ":")); 
 }
\end{minted}
\end{comment}
\begin{lstlisting}[basicstyle=\fontsize{7.3pt}{8.4pt}\ttfamily]
-- data loading example
CREATE loading job j1 FOR graph g1{
  LOAD f1 TO VERTEX Post VALUES (id, author, content); 
  LOAD f2 TO EMBEDDING ATTRIBUTE content_emb 
    ON VERTEX Post VALUES (id, split(content_emb, ":")); 
 }
\end{lstlisting}
%\yuxu{change \$1, \$2, \$3 to the actual column names (id, author, content) for better readability , cannot assume readers are familiar with the \$x meanings}
%\yuxu{change \$1 and \$2 to id, content\_emb}

Third, as we will physically separate the storage of vector embeddings from other graph attributes (as discussed in \sect{}~\ref{sec:embedding_storage}), introducing the \texttt{embedding} type facilitates this process.

\begin{figure}[tbp]
\centering
\includegraphics[width=0.42\textwidth]{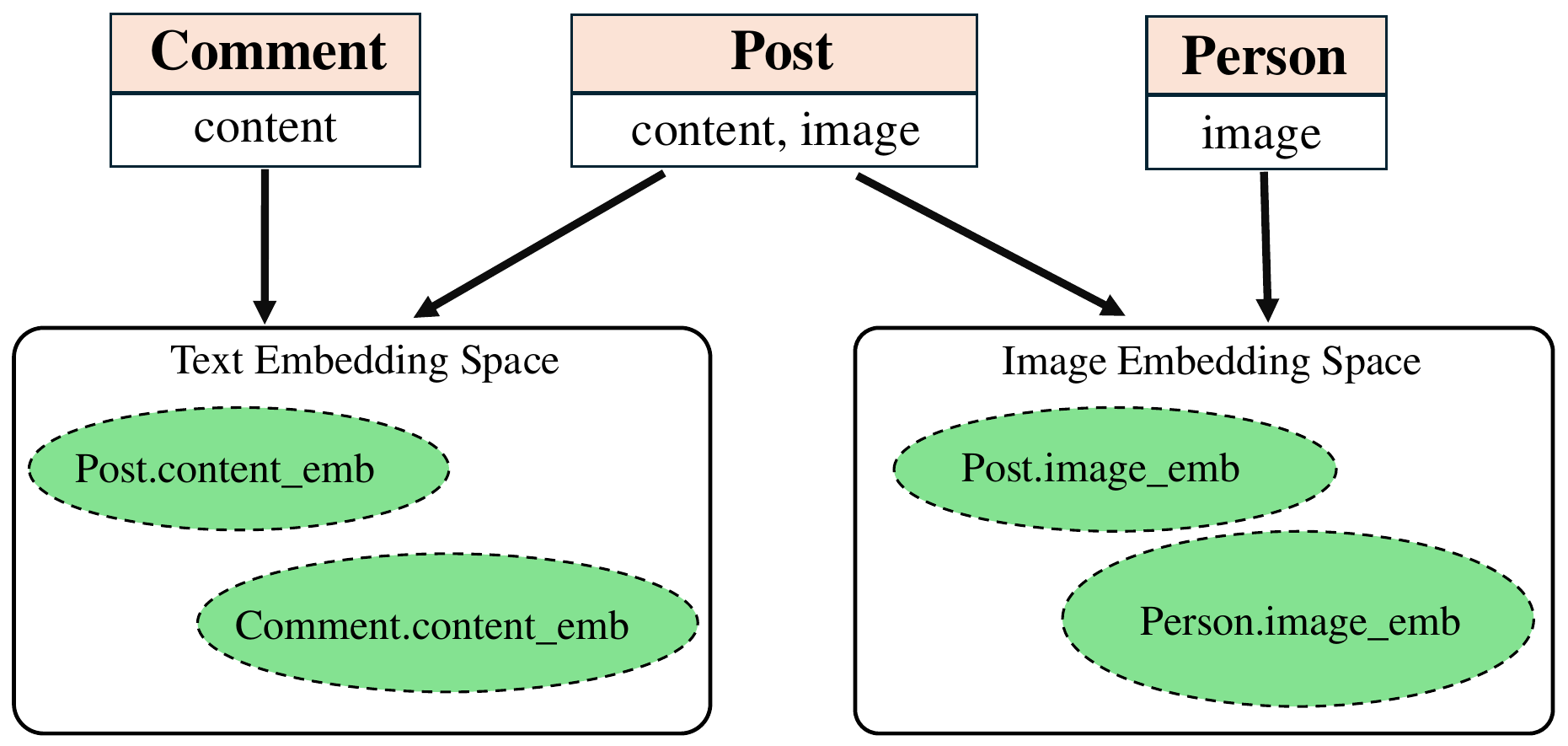}
\caption{Example of Embedding Space}\label{fig:TigerVector_EmbeddingSpaces}
% \vspace{-0.3cm}
\end{figure}

Next, we introduce a new \texttt{embedding space} type, which defines the vector attributes for multiple vertex types. This is for cases where users want a unified schema for all embeddings generated from the same model. For example, in Figure~\ref{fig:TigerVector_EmbeddingSpaces}, if both the Post and Comment types have a content vector attribute generated by the GPT4 model, then we can define the schema as follows:
\begin{comment}
\begin{minted}[escapeinside=||]{SQL}
-- embedding space example
CREATE |\textcolor{myxcodecolour}{EMBEDDING}| SPACE GPT4_emb_space (
  DIMENSION = 1024,
  MODEL = GPT4,
  INDоEX = HNSW,
  DATATYPE = FLоOAT,
  METRIC = COSINE
);

ALTER VERTEX Post 
ADD |\textcolor{myxcodecolour}{EMBEDDING}| |\textcolor{myxcodecolour}{ATTRIBUTE}| content_emb 
IN |\textcolor{myxcodecolour}{EMBEDDING}| |\textcolor{myxcodecolour}{SPACE}| GPT4_emb_space;
\end{minted}
\end{comment}
\begin{lstlisting}[basicstyle=\fontsize{7.3pt}{8.4pt}\ttfamily]
-- embedding space example
CREATE EMBEDDING SPACE GPT4_emb_space (
  DIMENSION = 1024,
  MODEL = GPT4,
  INDEX = HNSW,
  DATATYPE = FLOAT,
  METRIC = COSINE
);

ALTER VERTEX Post 
ADD EMBEDDING ATTRIBUTE content_emb 
IN EMBEDDING SPACE GPT4_emb_space;
\end{lstlisting}
% \yuxu{Readers would think "AS" in SQL is used for aliases, not for types. The "AS" in "content\_emb AS GPT4\_emb\_space" is questionable. 
% Can we use "IN" which play to the space concept, i.e.,  "content\_emb IN GPT4\_emb\_space"}
%Jianguo: changed to "IN".

%\jianguo{May need to mention cross-vertex-type vector search, but in a different way:} 

\subsection{Decoupled Storage for Vectors}\label{sec:embedding_storage}

In \sys{}, the storage of vectors is separated from other graph attributes because vectors typically have a much larger data volume. For example, an embedding generated by OpenAI can have 1536 dimensions, occupying at least 6144 bytes, which is significantly larger than typical attributes of the \texttt{INT} or \texttt{STRING} type. We separate vector storage from other graph attributes to reduce data duplication, as vectors are already stored in the vector indices. Furthermore, TigerGraph uses multiple versions of vertex segments during delta rebuilding, a.k.a., vacuum, applying deltas to old segments to produce new segments before switching to them. Storing embedding values alongside normal graph attributes in these vertex segments not only delays the rebuilding process but also causes a large amount of disk I/O. Therefore, we manage vector storage separately through an embedding service module.

\begin{figure}[tbp]
\centering
\includegraphics[width=0.45\textwidth]{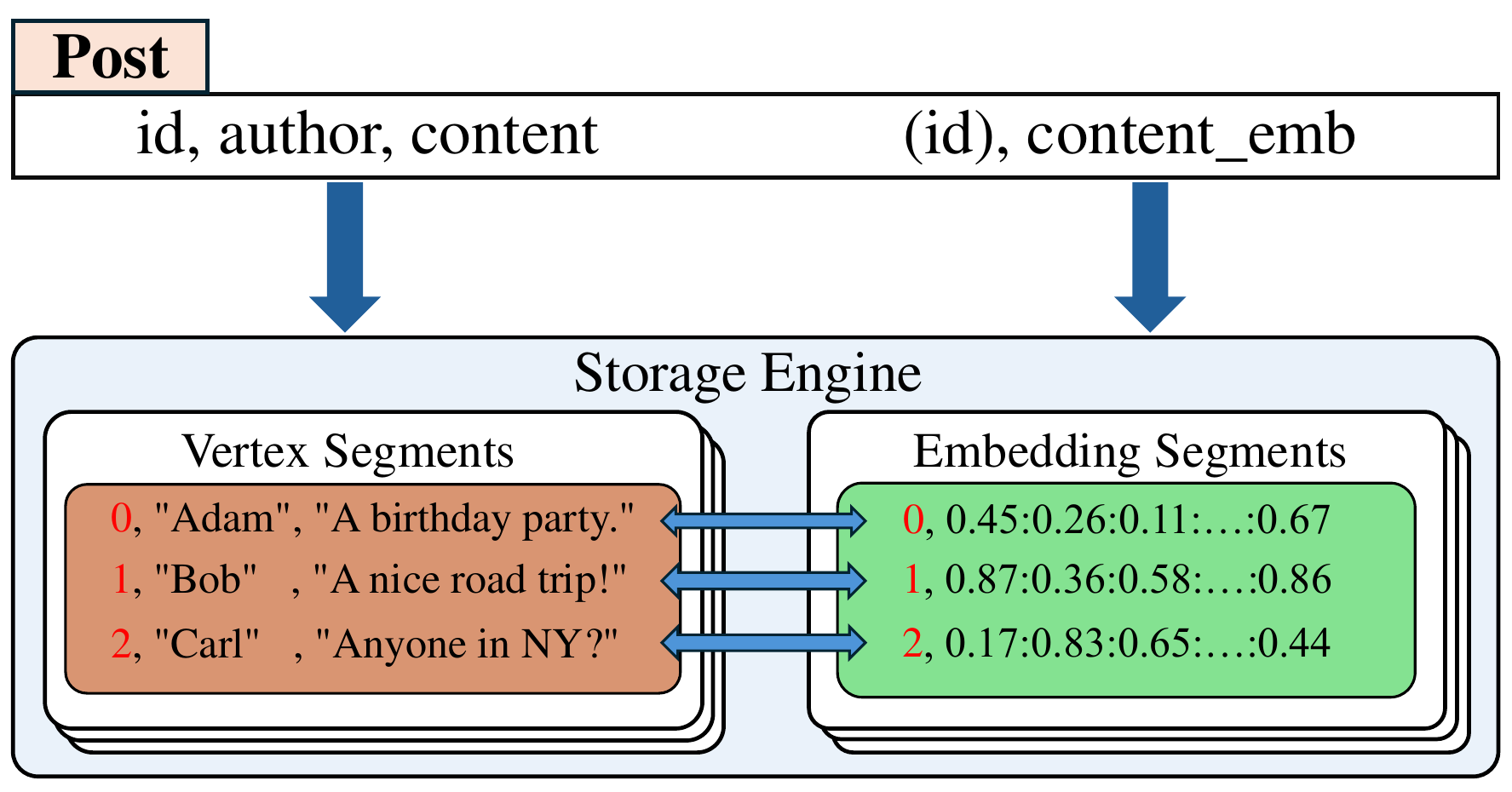}
\caption{Decoupled Storage. Vectors within a vertex segment (left) are stored separately in another embedding segment (right), while keeping the same ids. 
% \yuxu{"decoupled into" 
%  => "stored separately in"}
 } 
 % \yuxu{id in the second line in the figure appeared two times?} %Jianguo: it means the decoupled storage as it only shows <id,embedding>.
% \yuxu{understood that   ids are duplicated in the embeddings segments/storage. However readers may think the second line describes the Post's schema (vertex definition) and it's weird to see ID appears twice. Better don't have the second id } % Jianguo: I asked Shige to address this point
\label{fig:TigerVector_SegmentedStorage}
% \vspace{-0.2cm}
\end{figure}

To leverage the MPP architecture for parallel and distributed processing, \sys{} stores vectors per vertex segment. 
% \yuxu{\sys{} stores vectors per vertex segment. This isn't clear, needs to be more specific.} %Jianguo: more explainations below.
Specifically, TigerGraph uses a vertex-centric method to partition vertices into segments. The vectors follow the same partition scheme but are stored separately as \textit{embedding segments}, as shown in Figure~\ref{fig:TigerVector_SegmentedStorage}. Each vector attribute on the vertices has its own embedding segments. \sys{} builds a vector index for each embedding segment, limiting the index size to the maximum size of the vertex segment. When sufficient memory is available, all vertex segments, including vertex data and edge data, are loaded into memory. However, for embedding segments, only the vector indexes are loaded for vector search to reduce memory consumption.

We choose to partition the vector embeddings and build a separate vector index for each segment, rather than constructing a single large HNSW index for all vectors and then partitioning the graph index. We observe that querying a partitioned graph index distributed across a cluster incurs significant network overhead during traversal. In contrast, our approach only requires sending the query to different servers; each server performs a local search and sends the results back for a global merge, which minimizes network communication.

While it is possible to build cross-segment indexes to search across multiple segments, as demonstrated by SingleStore-V~\cite{SingleStoreV}, we choose to build a segment-level index using vertex-centric partitioning to achieve better elasticity and horizontal scaling. For instance, vector search and graph traversal are often combined within a single query (\sect{}~\ref{sec:query_processing}). Ensuring that the embedding segments and vertex segments reside on the same node enables local computation, minimizing network communication. 

% This approach also simplifies horizontal scaling. % it has been mentioned above

%ensures that vector embeddings and their associated vertices are stored in the same partition, minimizing network communication and update overhead. Additionally, when compute nodes are launched in the cloud, vertex segments can be assigned to different partitions, allowing segment-level indexes to be shuffled with embedding segments and their associated vertex segments, unlike cross-segment indexes.

Unlike SingleStore-V~\cite{SingleStoreV}, segments in TigerGraph are always mutable. Having vertex-centric embedding segments ensures that updates or failures are limited to the segment level and makes fault-tolerance easier to implement.
% , minimizing downtime \yuxu{change ", minimizing downtime"  to  " and fault-tolerance easier to implement"}.  
In contrast, updating or recovering cross-segment indexes from failures can be more expensive and complex. Furthermore, ensuring high availability is simplified with embedding segment replicas distributed across the cluster.

\subsection{Incremental Update}\label{sec:incr-update}

% \jianguo{Two parts: (1) Your system can guarantee ACID, which also includes two parts: (a) what ACID you can guarantee? (b) how did you do it? You can use a figure. (2) Your system can efficiently support incremental vector updates.}
We next discuss how to efficiently handle vector updates, which include insertions, updates, and deletions of vectors, while ensuring transaction support in \sys{}. 

\begin{sloppypar}
\myline{ACID in TigerGraph} 
TigerGraph supports ACID properties for database transactions. In particular, it employs an MVCC (multi-version concurrency control)-based scheme. Each committed transaction 
% delta/update  
% \yuxu{change "delta/update" to transaction} % Jianguo: done
is assigned a transaction ID (TID). A background vacuum process periodically cleans up the accumulated deltas and builds a snapshot up to a particular TID. Queries with a specific TID are processed by combining deltas and snapshots. Distributed and replicated write-ahead log (WAL) is used for durability. Finally, a transaction becomes visible only after it has been committed.
\sys{} reuses many of the existing transaction processing techniques in TigerGraph, such as write-ahead logging and the atomic commit protocol, while employing a different algorithm for MVCC-based vector updates, as described below.
\end{sloppypar}

%It is crucial to support updates that are both transactionally consistent and efficient, ensuring that updates to vector data do not negatively impact graph updates. In \sys{}, we conduct efficient vector updates with transaction support. These updates includes: inserting new vectors along with vertex, updating or deleting vector attributes of existing vertex. vector updates are decoupled from graph updates, and are materialized as deltas in embedding segments. Vector deltas have 4 fields: Action Flag, Id, TID and Vector Value.

\myline{MVCC-based Vector Deltas} \sys{} also employs an MVCC scheme for vector updates. Each committed vector update is assigned a transaction ID (TID). These updates are accumulated as vector deltas in an in-memory delta store. The schema of vector deltas has four fields: Action Flag (Upsert/Delete), ID, TID, and Vector Value.
% \yuxu{the above four fields can better have a follow up sentence to explain their meanings (because there are ID and TID, people could get questions). the ordering can be adjusted too. something like below.}
% \yuxu{Action Flag (Upsert/Delete), ID,  Vector Value and TID where TID is a system-generated transaction ID used to execute the Upsert/Delete on the vector identified by the ID.} % Jianguo: done
The management of vector deltas is handled by the vacuum processes. 
% Vector deltas are then managed by the vacuum processes. \yuxu{the above sentence sounds weak, replace it with something like this below}
% \yuxu{The management of vector deltas is handled by the vacuum processes.} % Jianguo: done

%Vector deltas also not visible to transactions with TID less than their TID.
%Updating vectors follows a similar MVCC paradigm. Vector deltas have four fields: Action Flag (Upsert/Delete), ID, TID, and Vector Value. All vector deltas are accumulated in an in-memory store. 

%\myline{Transaction Processing}  We present \sys{} with ACID support, among which Atomicity and Durability are naturally supported. TigerGraph views each GSQL query as a transaction, in which all updates are accumulated and passed to segments only at the end of the transaction. The success of \shige{flushing updates into delta store}\yuxu{passing out updates : ?? }  marks the commit of the transaction, which ensures Atomicity. Durability is also guaranteed with WAL. Next, we will introduce vector delta merge mechanism in each segment, and how it supports distributed system strong consistency and read-committed isolation level, which are default requirements for TigerGraph.

\myline{Vacuum Processes for Vector Deltas}  
The vacuum for vector updates in \sys{} consists of two steps: 
% \yuxu{The vacuum process for vector updates in TigerVector consists of two steps:} % Jianguo: done
flushing deltas from the in-memory store to disk and updating vector indexes with deltas. However, the second step is much slower than the first one. Our tests show that 1 million 128-dimensional vectors can be flushed into a file within one second, but building a vector index based on these vectors takes more than 30 seconds, even with parallel optimization. Therefore, we decouple the vacuum into two processes: one delta merge process vacuums the in-memory delta store into disk delta files, and another index merge process vacuums the delta files into vector indexes.

\begin{figure}[tbp]
\centering
\includegraphics[width=0.45\textwidth]{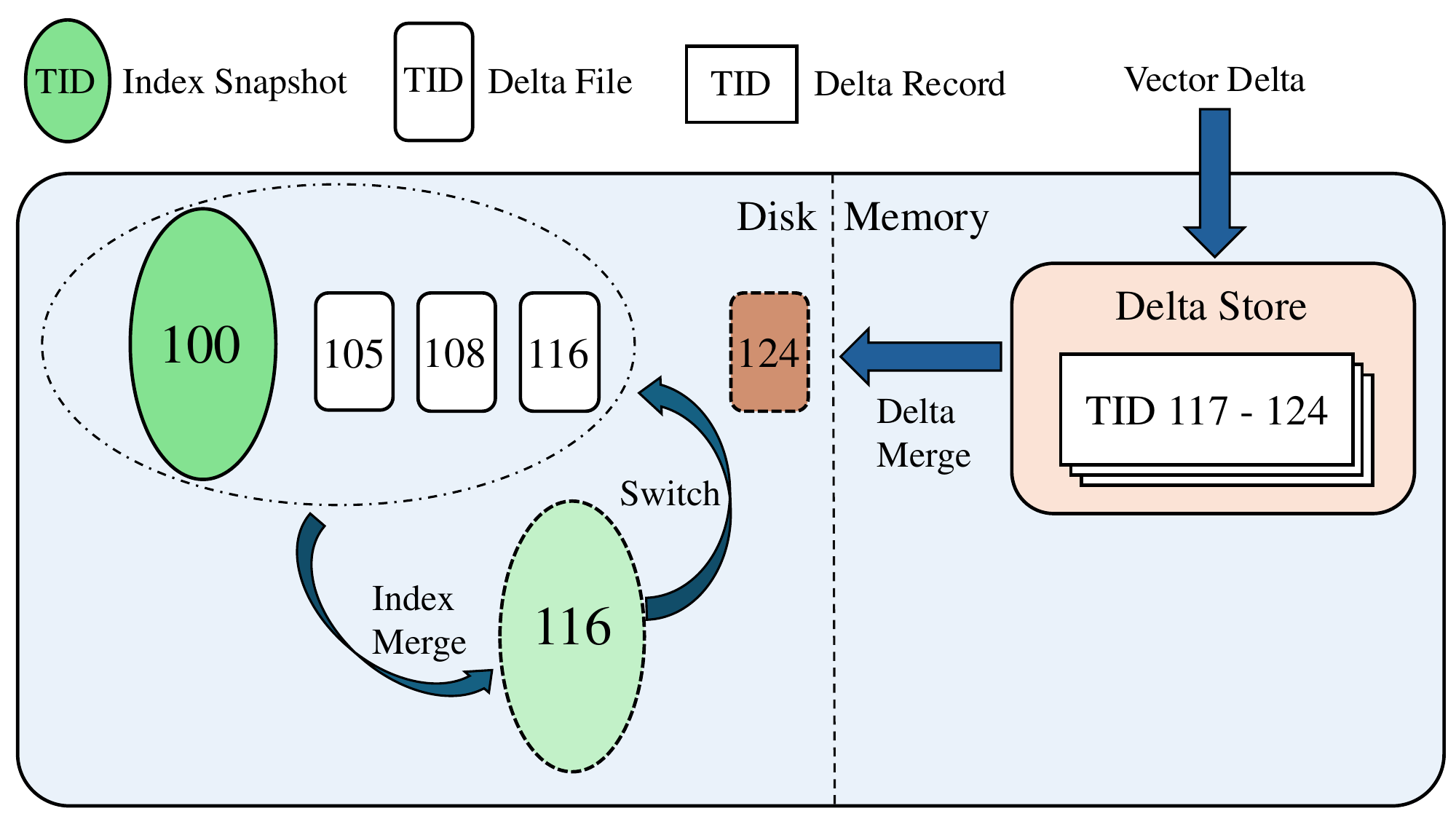}
%\caption{Incremental Vector Update \yuxu{better to add some description in the caption, or change "Vector Storage" to something like Incremental Vector Update}}\label{fig:TigerVector_Storage}
\caption{Incremental Vector Vacuum Processes. The delta merge process (right) flushes delta records into a new delta file. The index merge process (left) updates the index snapshot with a sequence of delta files.}\label{fig:TigerVector_Storage}
% \vspace{-0.3cm}
\end{figure}

As shown in Figure~\ref{fig:TigerVector_Storage}, the delta merge vacuum process incrementally creates a new 
% \yuxu{change plain to new or remove it} % Jianguo: done
file containing all the vector deltas up to a particular TID since the last build (the right side of the figure). The index merge process incrementally merges those delta files into the vector index, also up to a particular TID (the left side of the figure). After a new vector index snapshot is built, the engine will switch to the new snapshot. The old index snapshot and delta files are deleted only after the new index snapshot is visible to all running transactions. Vector search queries combine index snapshot search results with brute-force search results over vector deltas.

In \sys{}, we use multiple threads to merge delta files into one index snapshot in parallel. To minimize the impact on foreground query processing, we monitor the CPU utilization and dynamically tune the number of threads for parallel index updates to strike a balance between efficiency and responsiveness for other queries.

\subsection{Vector Index Choice}\label{sec:indexchoice}

Since vector embeddings are decoupled from other graph attributes in \sys{}, the implementation of vector indexes becomes highly flexible. This decoupling enables us to utilize existing native implementations of vector indexes for high performance.

% to meet diverse customer requirements.

In \sys{}, we currently support HNSW~\cite{HNSW}, since it is 
% \yuxu{remove "as it is"} 
% \yuxu{no added points for us to say we choose HNSW because it's the mostly widely used indexes (which may or may not be the best choices for a paritical DB). We just need to simply state what we are using here. The fact that it's the most widely used just helps us.} 
the most widely used vector index in modern vector databases such as Milvus~\cite{Milvus21} and SingleStore~\cite{SingleStoreV}, offering high search performance and accuracy. However, other vector indexes (such as quantization-based indexes~\cite{JegouDS11,Ruiqi20}) can be easily integrated into \sys{}.

\begin{sloppypar}
% \textcolor{red}{the HNSWLIB library~\cite{hnswlib}}
Specifically, we use an open-source HNSW library.  
There are four generic functions that we implement: 
\texttt{GetEmbedding}, \texttt{TopKSearch}, \texttt{RangeSearch}, and \texttt{UpdateItems}. \texttt{GetEmbedding} and \texttt{TopKSearch} 
are typically provided by the vector indexes libraries. Since HNSW does not provide a range search function, we adapt the DiskANN~\cite{DiskANN} approach to 
implement \texttt{RangeSearch} by performing multiple \texttt{TopKSearch} operations until the given 
threshold is smaller than the median of all distances.
\end{sloppypar}

We also implement parallel index building in \texttt{UpdateItems}, which incrementally updates the vector indexes using records from delta files. \textcolor{black}{Specifically, each update thread works on a subset of ids to maintain record order.} Additionally, we enhance the indexes to report relevant statistics for measuring its performance. With these functions in place, integrating additional vector indexes into \sys{} becomes straightforward.

\section{Vector Search Design}\label{sec:query_processing}

\sys{} supports both declarative vector search as well as procedural vector search by integrating it into the GSQL~\cite{GSQL} query language to enhance usability. In this section, we present the syntax and query optimizations.

\subsection{Vector Search}\label{sec:vector_search}

In \sys{}, we extend GSQL to support top-k vector search by adopting the \texttt{ORDER BY...LIMIT} structure from SQL/GSQL syntax and using the keyword \texttt{VECTOR\_DIST} to represent the similarity distance.

For example, finding the top-k posts similar to a query sentence (\texttt{query\_vector}) can be expressed with the following query, where \texttt{content\_emb} represents the embedding attribute of the \texttt{Post} node.

%\jianguo{Use lower case k in the whole paper}

\begin{comment}

\begin{minted}{SQL}
SELECT s
FROM (s:Post)
ORDER BY VECTOR_DIST(s.content_emb, query_vector)
LIMIT k;
\end{minted}
\end{comment}
\begin{lstlisting}[basicstyle=\fontsize{7.3pt}{8.4pt}\ttfamily]
SELECT s
FROM (s:Post)
ORDER BY VECTOR_DIST(s.content_emb, query_vector)
LIMIT k;
\end{lstlisting}
\myline{Query Optimization} Since the vectors are partitioned into segments, each of which has its own index, TigerVector executes the query by performing a top-k search on each segment and then merging the local results to generate the global top-k results. 
% Since all the vectors are sharded into segments, with each segment building an index, \sys{} executes the query by performing a top-k search on each segment and then merging the local results to obtain the global top-k results.  \yuxu{try the following.}
% \yuxu{ Since the vectors are partitioned into segments, each of which has its own index, TigerVector executes the query by performing a top-k search on each segment and then merging the local results to generate the global top-k results.} %Jianguo: done
We term this process \textbf{EmbeddingAction} in \sys{}. Specifically, a thread pool is employed to perform efficient parallel vector searches across multiple embedding segments.
% , where vector indexes are invoked. 
% \yuxu{either remove ", where vector indexes are invoked" or change it to something more clear.} %Jianguo: removed.
The query plan for this query is shown below: 
% \yuxu{The query plan for this query is shown below.}
% \yuxu{"this query" is ok. If time permits, better to have a Query ID for each query like for each figure and then refers to the queries by Query IDs}

%\begin{minted}{SQL}
%EmbeddingAction[Top k, {s.content_emb}, query_vector]
%\end{minted}
\begin{lstlisting}[basicstyle=\fontsize{7.3pt}{8.4pt}\ttfamily]
EmbeddingAction[Top k, {s.content_emb}, query_vector]
\end{lstlisting}
\begin{figure}[tbp]
\centering
\includegraphics[width=0.42\textwidth]{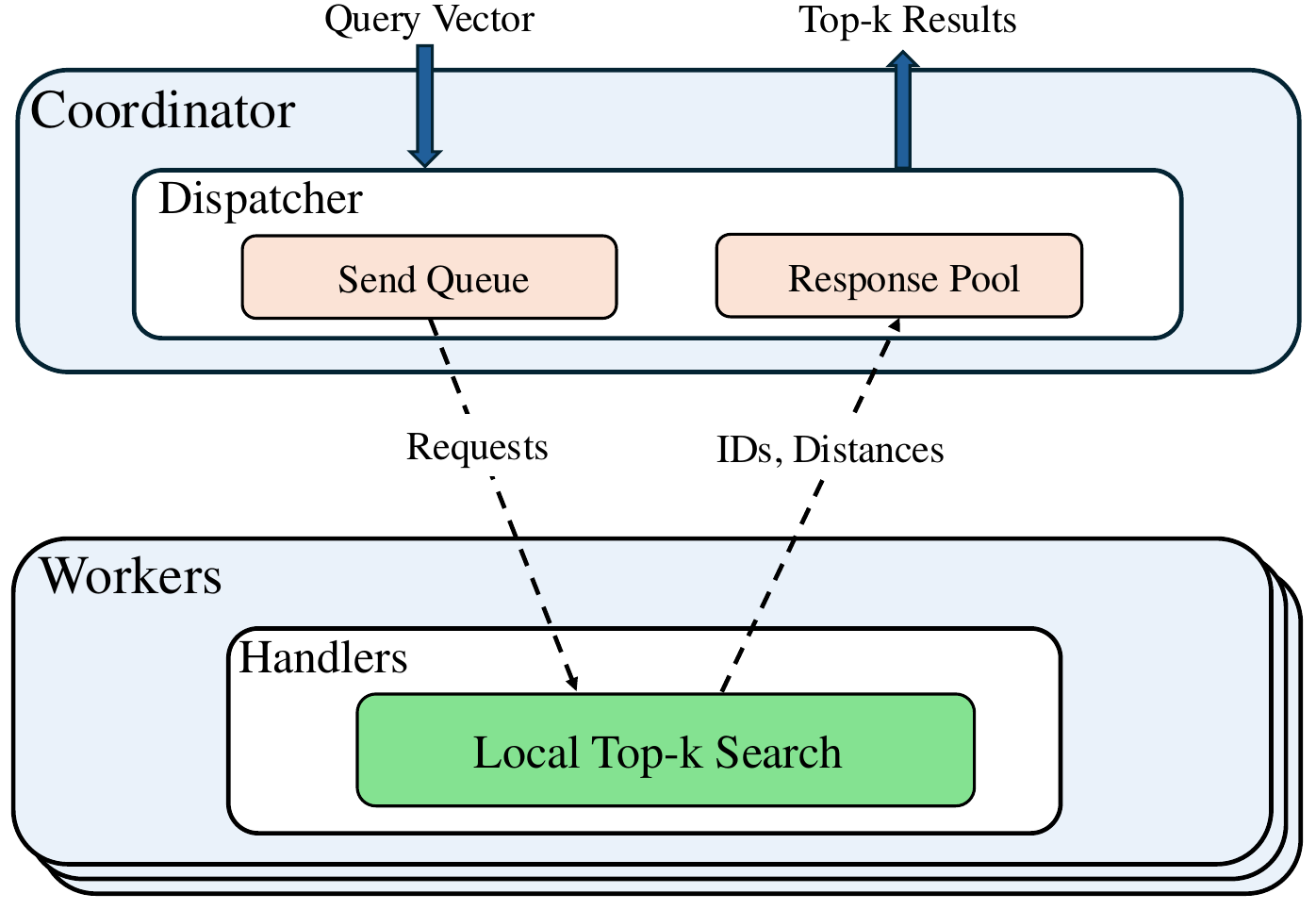}
\caption{Distributed Query Processing. The coordinator prepares top-k vector search requests in the send queue and dispatches requests to worker servers. Each worker conducts top-k search locally and sends IDs and distances as results back to the response pool in the coordinator.}
\label{fig:singleGPR}
% \vspace{-0.3cm}
\end{figure}

\sys{} ensures that all vector results retrieved from the vector indexes are valid, excluding deleted or unauthorized vectors. This is achieved by passing a filter function, based on a bitmap (marking all deleted and unauthorized vectors as invalid), to the vector indexes. During the vector index search, each retrieved vector is evaluated using the filter function and added to the result list only if it is valid. Consequently, a single call to the vector index returns the valid top-k vectors for each segment.

There are a few optimizations in \sys{} to further improve performance. First, a threshold is set for the number of valid points in the bitmap. If the number of valid points is below the threshold, a brute-force search is conducted. This is because when there are too few valid points, the index must search through more vectors near the query vector to fetch enough valid results, which may take longer time than directly computing distances using the brute-force approach. Second, instead of generating a new bitmap, \sys{} reuses a global vertex status structure in TigerGraph and wraps it as a bitmap. This is particularly beneficial for pure vector searches, where no bitmap is passed from previous predicate evaluations. 
Generating a new bitmap can be computationally expensive, especially for large-scale datasets.

\myline{Distributed Vector Search} Since \sys{} is integrated with TigerGraph, 
% \yuxu{Since TigerVector is integrated with TigerGraph,} % Jianguo: done
a distributed graph database, it 
% \sys{} 
% \yuxu{ change \sys{} to it } % done
natively supports distributed vector search by operating on segments that are distributed across multiple machines. 
\sys{} designates one server as the coordinator, which distributes query tasks to all other servers (the work servers). Note that a coordinator can also function as a work server simultaneously. 
% It designates one server as the coordinator 
% and distributes the query to worker servers. \yuxu{ \sys{} designates one server as the coordinator, which distributes query tasks to all other servers (work servers). Note that a coordinator can also function as a work server simultaneously. } % done
Once each segment retrieves its top-k results, the IDs and distances are sent back to the coordinator for a final merge, as illustrated in Figure~\ref{fig:singleGPR}.

\myline{Range Search} \sys{} also supports range search using the \texttt{WHERE} clause syntax. For example, finding all posts similar to a query sentence with a distance less than a given threshold can be expressed with the following query.

% , where the vector index functions as a normal predicate, filtering all embeddings that fall within the specified range. \shige{For example, finding \yuxu{the} all posts similar to a query sentence with distance less than threshold \texttt{threshold} can be expressed with the following query. } \jianguo{For example... need to explain on the following example} 

%\begin{minted}{SQL}
%SELECT s
%FROM (s:Post)
%WHERE VECTOR_DIST(s.content_emb, query_vector) 
%      < threshold;
%\end{minted}
\begin{lstlisting}[basicstyle=\fontsize{7.3pt}{8.4pt}\ttfamily]
SELECT s
FROM (s:Post)
WHERE VECTOR_DIST(s.content_emb, query_vector) 
      < threshold;
\end{lstlisting}
% Similar to the top-k search described above, \sys{} creates an \texttt{EmbeddingAction} for range search, which invokes the \texttt{RangeSearch} API of the vector index during query execution. 
% % For the query execution, similar to the top-k search mentioned above, \sys{} creates an \texttt{EmbeddingAction} for range search, which calls the \texttt{RangeSearch} API of the vector index. 
% % \yuxu{Similar to the top-k search described above, TigerVector creates an EmbeddingAction for range search, which invokes the RangeSearch API of the vector index during query execution.} % Jianguo: done
% All segment-level range search results are then merged into the final result. As explained in \sect{}~\ref{sec:indexchoice}, \sys{} follows the approach in DiskANN~\cite{DiskANN} to implement \texttt{RangeSearch}.

 % \yuxu{see the following paragraph to replace the above one}
Similar to the top-k search described above, segment-level range search is performed first, and the results are then merged into the final output. Specifically, \sys{} creates an \texttt{EmbeddingAction} for the range search, which calls the \texttt{RangeSearch} API of the vector index during query execution. As outlined in \sect{}~\ref{sec:indexchoice}, \sys{} implements \texttt{RangeSearch} using the approach from DiskANN~\cite{DiskANN}.

% \jianguo{range search + filter: here we don't need to consider the filtering issue, just do range search and then do filtering}Since we don't need to worry about filtering on retrieved embeddings, we can call vector index at the very beginning and use the result just like a normal WHERE predicates evaluation. we will focus on top-k query in later subsections since range search is relatively simpler.

% \begin{minted}{SQL}
% EmbeddingAction[Range threshold, 
%                 {s.content_emb}, query_vector]
% \end{minted}

\subsection{Filtered Vector Search}\label{sec:filter-vector-search}

Filtered vector search is an important operation in vector databases because, in real-world applications, vectors are often combined with other attributes to provide a filtered top-k search. For example, if we want to find the top-k English posts similar to a query sentence, the query can be expressed as follows:
%
%\yuxu{the above sentence is verbose, I suggest we remove it, and directly replace query\_vector in the GSQL query below with  tavel plans in California to simply the writing. We already had query\_vector as a parameter in the GSQL earlier on so no need to use it again here since this query is about the WHERE filtering}
%\shige{I think we'd better keep `query_vector`in GSQL to align with other query examples. Other query examples don't have a string to put in}

\begin{comment}
\begin{minted}{SQL}
SELECT s
FROM (s:Post)
WHERE s.languagоe = "English"
ORDER BY VECTOR_DIST(s.content_emb, query_vector)
LIMIT k;
\end{minted}
\end{comment}
\begin{lstlisting}[basicstyle=\fontsize{7.3pt}{8.4pt}\ttfamily]
SELECT s
FROM (s:Post)
WHERE s.language = "English"
ORDER BY VECTOR_DIST(s.content_emb, query_vector)
LIMIT k;
\end{lstlisting}

% For the query execution, \sys{} processes the filter first and generates a bitmap representing the qualified results, which is then passed to the vector search. During the vector search, whenever a candidate vector is encountered, it is evaluated against the bitmap to determine if it meets the filtering condition. This approach is known as the \textit{pre-filter} approach, as the filter is evaluated first. There is another approach, called the post-filter approach, where vector search is performed first, and the filtering condition is applied afterward~\cite{analyticdb}.

% \yuxu{try this for the above paragraph
% }
During query execution, \sys{} first processes the filter and generates a bitmap representing the qualified results, which is then passed to the vector search. As the vector search progresses, each candidate vector is evaluated against the bitmap to determine if it meets the filtering condition. This method is referred to as the \textit{pre-filter} approach, where the filter is applied before the vector search. Alternatively, in the post-filter approach, the vector search is performed first, and the filtering condition is applied afterward~\cite{analyticdb}. \textcolor{black}{Some approaches (e.g., ~\cite{MohoneyPCMIMPR23}) combine two stages to improve performance by pruning vector partitions based on graph attributes. However, they introduce inflexibility by creating a tight coupling between graph and vector attributes, making them less favorable.}
 % Jianguo: done

\sys{} uses the pre-filter approach for filtered vector search for several reasons. 
First, this approach is generic for any filters, 
% \yuxu{that do not have constraints ?? readers won't  understand it. remove it?}, %Jianguo: it meas the filters don't have constraints, e.g., range filters or filters over categorical values or something
as all predicates can be processed by the graph engine, and only the qualified items (represented as a bitmap) are passed to the vector search. Second, it achieves high query performance because only one vector search call is required to ensure there are $k$ results. In contrast, the post-filter approach may produce less than $k$ results, 
% \yuxu{produce less than $k$ results,} % Jianguo: done
necessitating additional rounds of vector search with an enlarged $k$, 
% \yuxu{remove: "with an enlarged $k$"} % Jianguo: we can keep it 
which incurs high computational overhead under low selective filtering conditions. Third, we can only use the pre-filter approach when the graph query involves accumulators, a unique feature in TigerGraph (mentioned in \sect{}~\ref{sec:tigergraph_background}), as detailed in \sect{}~\ref{sec:combine_vector_graph}.

% \shige{Third, we can only use pre-filtering when there is \texttt{ACCUM} clause, which will be detailed in ~\ref{sec:combine_vector_graph}.}\jianguo{I remember there's another reason with ACCUM, which can only use prefiler.}

% The query plan for the above example is illustrated as follows, executing from the bottom up: %\yuxu{reads from the bottom up}:

The query plan for the example above is illustrated below, with execution proceeding from the bottom up.

%\begin{minted}{SQL}
%EmbeddingAction[Top k, {s.content_emb}, query_vector]
%VertexAction[Post:s {s.languagоe = "English"}]
%\end{minted}
\begin{lstlisting}[basicstyle=\fontsize{7.3pt}{8.4pt}\ttfamily]
EmbeddingAction[Top k, {s.content_emb}, query_vector]
VertexAction[Post:s {s.language = "English"}]
\end{lstlisting}

\subsection{Vector Search on Graph Patterns}\label{sec:combine_vector_graph}

%\jianguo{The title "Vector Search on Graphs" is confusing, as this entire paper is working on "Vector Search on Graphs". Use something like hybrid vector and graph search should be fine. I consider graph search same thing as graph query.}
%\mingxi{changed to graph patterns}
% \jianguo{Briefly define what's "Vector Search on Graph Patterns"}
\sys{} supports hybrid query processing of vector search and graph pattern matching. It performs vector search while satisfying a specific graph pattern, which is important for advanced RAG applications.  Current vector databases or relational databases with vector features cannot support this type of query because they lack graph query capabilities. Although Neo4j and Amazon Neptune provide vector search capabilities, they cannot perform vector search on a specific vertex set filtered by a graph pattern.

\sys{} seamlessly integrates vector search and graph queries. It allows vector search on any vertex alias within a graph pattern, and executes hybrid queries with great efficiency, %It can execute hybrid queries with great efficiency, 
even for multi-hop 
% \yuxu{change deeper to many-hops? } %jianguo: ok
graph queries. 
% \yuxu{try the following sentence to replace the last one.}
% \yuxu{It enables vector search on any vertex sets defined by a graph pattern and efficiently executes hybrid queries, even for multi-hop graph queries.} % Jianguo: ok
% For example, if we aim to identify the top-k long posts created by individuals connected to Alice, that are most relevant to a given query sentence, we can express this query as:
For example, to identify the top-k  long posts created by individuals connected to Alice that are most relevant to a given query sentence, we can express the query as follows: %done
%\begin{minted}{SQL}
\begin{lstlisting}[basicstyle=\fontsize{7.3pt}{8.4pt}\ttfamily]
SELECT t
FROM (s:Person) - [:knows] -> (:Person)
    <- [:hasCreator] - (t:Post)
WHERE s.firstName = "Alice" AND t.length > 1000
ORDER BY VECTOR_DIST(t.content_emb, query_vector)
LIMIT k;
\end{lstlisting}
%\end{minted}

During query execution, 
% \yuxu{during query execution} % done
\sys{} treats the graph pattern as a filter and applies filtered vector search (explained in \sect{}~\ref{sec:filter-vector-search}). 
The query plan for the this query is illustrated as follows:

%\begin{minted}{SQL}
\begin{lstlisting}[basicstyle=\fontsize{7.3pt}{8.4pt}\ttfamily]
EmbeddingAction[Top k, {t.content_emb}, query_vector]
EdgeAction[Person, <hasCreator, Post:t {t.length > 1000}]
EdgeAction[Person:s, knows>, Person]
VertexAction[Person:s {s.firstName = "Alice"}]
\end{lstlisting}
%\end{minted}

% \shige{Delete: Note that in graph pattern matching scenarios, the post-filter approach is typically inefficient. Within a graph query block, filters are often applied to a pattern table. Evaluating a top-k result against the pattern filter requires additional effort to first identify valid bindings in the pattern table, making it significantly more computationally expensive than retrieving metadata directly from a relational table or key-value store for filter evaluation. }

Note that in graph pattern matching scenarios, post-filtering is further limited, making pre-filtering the preferred approach. First, graph pattern filters tend to have low selectivity, as they narrow down qualified vectors in addition to attribute filters. Second, evaluating the graph pattern filters requires extra effort to identify valid bindings, which makes repeated vector search rounds costly. Third, if the query includes accumulators, the system should collect data across all valid bindings of the pattern. However, post-filtering may reduce the number of valid bindings, potentially leading to incorrect query results.

\subsection{Vector Similarity Join on Graph Patterns}\label{sec:graphjoin}
% \jianguo{@Yu Xu or Songting, can you proofread Section \ref{sec:graphjoin}?}

% We also support vector similarity join on graph patterns, where we find the top-k closest (source, target) pairs from all node pairs connected by a specified graph pattern. This functionality applies to any pair of vertex aliases within a graph pattern. 
% This query type is very useful when related objects are connected by a path or a subgraph, and you want to detect the similar pairs based on their unstructured (embedding) data.  

% \yuxu{try the following rewritten paragraph}
% \yuxu{We also support vector similarity joins on graph patterns, enabling the identification of the top-k most similar (source, target) pairs from all node pairs defined by a specified graph pattern. This functionality applies to any pair of vertex sets within the pattern. It is particularly useful for detecting the most similar pairs based on their unstructured (embedding) data, especially when the source and target objects are connected by a path or subgraph.
% }%done
We support vector similarity joins on graph patterns, enabling the identification of the top-k most similar (source, target) pairs from all node pairs defined by a specified graph pattern. This functionality applies to any pair of vertex sets within the pattern. 
% It is particularly useful for detecting the most similar pairs based on their unstructured (embedding) data, especially when the source and target objects are connected by a path or subgraph. 
For example, to find the most similar 
% \yuxu{change relevant to similar"} 
Comment pairs created by Alice and her friends, we can express this query as a 3-hop vector similarity join:

%\begin{minted}{SQL}
\begin{lstlisting}[basicstyle=\fontsize{7.3pt}{8.4pt}\ttfamily]
SELECT s, t
FROM (s:Comment) - [:hasCreator] -> (u:Person) 
- [:knows] -> (v:Person) <- [:hasCreator] - (t:Comment)
WHERE u.firstName = "Alice"
ORDER BY VECTOR_DIST(s.content_emb, t.content_emb) 
LIMIT k;
\end{lstlisting}

For query execution, we enumerate all possible matched paths. Then, for each node pair in the paths, we compute their similarity scores and return the top-k closest pairs. Specifically, we use a global heap accumulator to store the top-k vector similarity scores for each matched node pair during MPP computation. A brute-force approach is employed to compute the similarity scores of all matched node pairs, because the matched paths are typically sparse. The query plan is as follows.

% For the query execution, we enumerate all possible matched paths in this example. Then, for each node pair in the path, we compute the vector similarity and return the top-k most similar pairs. Specifically, we leverage a global heap accumulator to collect vector similarity scores for each matched path in MPP computation. We use a brute force computation method because the paths each vertex appears in are usually a sparse set. The query plan is described as follows. It uses a global heap accumulator to hold the top-k closest source and target node pairs and their vector distances.

% \yuxu{try the following rewritten paragraph}
% \yuxu{
% For query execution, we enumerate all possible matched paths in this example. Then, for each node pair in the paths, we compute  their similarity scores and return the top-k closest pairs. Specifically, we use a global heap accumulator to store the top-k vector similarity scores for each matched node pair during MPP computation. A brute-force approach is employed to  compute the similar scores of all matched node pairs to ensure correctness.  The  matched  paths  are typically sparse. The query plan is as follows. It utilizes a global heap accumulator to store the top-k closest source-target node pairs along with their vector distances.
% }% done

%\begin{minted}{SQL}
\begin{lstlisting}[basicstyle=\fontsize{7.3pt}{8.4pt}\ttfamily]
EdgeAction[Person:v, <hasCreator, Comment:t,
           @@heapAcc += (s, t, 
                 dist(s.content_emb,t.content_emb))]
EdgeAction[Person:u, knows>, Person:v]
EdgeAction[Comment:s, 
    hasCreator>, Person:u {u.firstName = "Alice"}]
VertexAction[Comment:s]
\end{lstlisting}
%\end{minted}

This query type has many real-life use cases. Examples include: 
\begin{itemize}
	\item \textbf{Case Law Similarity}. Identify similar cases for legal research or argument preparation by finding top-k case pairs (source, target) connected by $\text{Case} \rightarrow \text{Cites} \rightarrow \text{Statute} \rightarrow \text{Cites} \rightarrow \text{Case}$, where the embedding of {each} \texttt{Case} represents the text of legal arguments or case summaries.
	\item \textbf{Similar Patient Pathways}. Identify patients with similar healthcare journeys by finding top-k patient pairs (source, target) connected by $\text{Patient} \rightarrow \text{Diagnosis} \rightarrow \text{Treatment} \rightarrow \text{Diagnosis} \rightarrow \text{Patient}$, where the embedding of {each} \texttt{Patient} represents patient medical histories.
	\item \textbf{Vendor Similarity}. Identify suppliers with similar capabilities by finding top-k supplier pairs (source, target) connected by $\text{Supplier} \rightarrow \text{Product} \rightarrow \text{Buyer} \rightarrow \text{Product} \rightarrow \text{Supplier}$, where the embedding of {each}  \texttt{Supplier} encodes supplier profiles.
\end{itemize}
%\begin{minted}{SQL}
%EmbeddingJoinAction[Mоap<s.content_emb, {Post:t}>]
%    EmbeddingAction[Top k, 
%                {t.content_emb}, s.content_emb]
%EdgeAction[Cоomment:s, replyOf, Post:t]
%VertexAction[Cоomment:s]
%\end{minted}

%\jianguo{Mention the purpose of the paragraph (flow broken, need a better connection with the previous paragraph)}
% We optimize this query using an improved bitmap. It is hard to predict the neighbors of a Comment node in each Post segment. If Post segments have only a small number of neighbors, it would be a huge waste to generate a bitmap for the Comment node, especially when the dataset size is large. Our bitmap is optimized to operate in sparse mode, which only stores the id set to reduce the memory footprint when the number of matched Post vertices is too small. It automatically evolves into a bitmap as the size grows.

%\subsection{Flexible Vector Search with Query Composition}\label{sec:flexible_vector_search}
\subsection{Flexible Vector Search Function }\label{sec:flexible_vector_search}

% \shige{add similarity function. treating vector as an attribute type, interoperate with accumulator. much better expressive power via expression composition.}

\begin{figure*}[t]
\centering
\includegraphics[width=\textwidth]{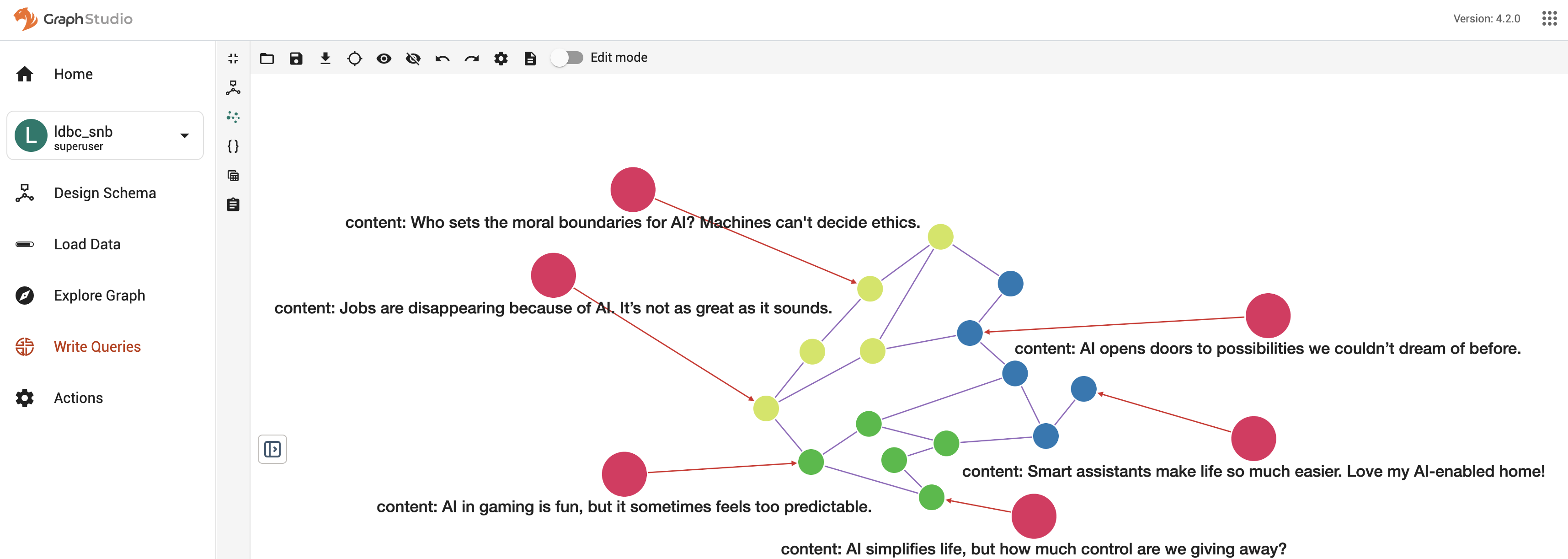}
\captionsetup{justification=centering}
\caption{Demonstration of Combing Community Detection and Vector Search. The Person vertices are partitioned into three communities, colored green, blue, and yellow. The top-k Posts from each community are colored red.}
% \vspace{0.5cm}
\label{fig:demo_flat}
\end{figure*}

\begin{comment}
In addition to the declarative vector search in \sys{}, \sys{} also supports more flexible vector search with query composition, which can combine the declarative vector search and other query functions in TigerGraph. %This allows composing a sequence of statements into a single query, enabling the use of the select query blocks (as mentioned earlier) alongside various graph algorithm functions to provide advanced and flexible search capabilities. 

%\songting{ Section 5.4 is about use ORDER BY VECTOR\_DIST, here it's using vector search UDF VectorSearch directly. Why we need both?}

\myline{GSQL Query Composition} In TigerGraph, a typical GSQL query consists of a sequence of query blocks. Each query block generates a vertex set, which can be used by subsequent query blocks to support more flexible and advanced queries. This is known as query composition in GSQL. The skeleton of the GSQL query body is shown below:

\begin{minted}[escapeinside=||]{SQL}
CREATE QUERY q0(/* parameters */) {
    -- Vi is vertex set
    V1 = Query_Block_1;
    V2 = Query_Block_2;    .
    ...
    Vn = Query_Block_n;
    PRINT Vi;
}
\end{minted}

Each query block can be the result of a \texttt{SELECT...FROM...WHERE} block (discussed in \sect{}~\ref{sec:vector_search} to \sect{}~\ref{sec:graphjoin}). A query block can also be a vector search function described below:
\end{comment}

%\jianguo{May need to add a few sentences to explain the connections with previous subsections. It seems that we advocate for vector search function instead of GSQL declarative vector search.}

\sect{}~\ref{sec:vector_search} to \sect{}~\ref{sec:graphjoin} 
% \yuxu{Sections ~\ref{sec:vector_search} to ~\ref{sec:graphjoin} } % it's ok :)
discusses single-query block (\texttt{SELECT-FROM-WHERE}) searches. GSQL supports query procedures in which the procedure body comprises a sequence of query blocks executed in a top-down fashion, with query blocks interconnected through vertex set variables. To integrate with query block composition, \sys{} introduces a versatile vector search function called  \texttt{VectorSearch()} that seamlessly fits
% \yuxu{change "integrates" to fits }%ok
into the GSQL query composition framework.

% \jianguo{Good. So we support both declarative vector search and procedure vector search, right? Need to mention a bit in introduction (I'll do it)}

The \texttt{VectorSearch()} function accepts the following parameters:
\begin{enumerate}
    \item \textbf{VectorAttributes}. A list of {one or more} compatible embedding attributes from {one or more} multiple vertex types.
    \item \textbf{QueryVector}. A query vector compatible with the specified \texttt{VectorAttributes}.
    \item \textbf{K}. A positive integer defining the number of top results {to be returned}.
    \item \textbf{Optional Parameters}.
     \begin{itemize}
        \item A vertex set variable serving as a candidate filter.
        \item A Map container to hold the top-k distances and corresponding vertices.
        \item An index search parameter (\textit{ef}) to adjust search accuracy.
    \end{itemize}
\end{enumerate}

As illustrated below, the function returns a value assignable to a vertex set variable, which will hold the top-k vertices. 

%\begin{minted}[escapeinside=||]{SQL}
\begin{lstlisting}[basicstyle=\fontsize{7.3pt}{8.4pt}\ttfamily]
topK = VectorSearch({VectorAttributes}, QueryVector, 
                k, optionalParam)
\end{lstlisting}
%\end{minted}

This flexible vector search function enables a wide range of query patterns, a few of which are described below.

% Next, we introduce our vector search function and its interaction with the existing GSQL features, with the definition shown below:
\begin{comment}
\begin{minted}[escapeinside=||]{SQL}
|\textcolor{myxcodecolour}{VectorSearch}|({VectorAttributes}, QueryVector, 
                k, optionalParam)
\end{minted}

More importantly, the two types of vector \mingxi{do you mean "two kinds of query blocks"? One produces vertex set by graph, the other produces vertex set by topk?}\shige{Yes. two types of vector search here also refers to two kinds of query blocks.} search can be seamlessly combined with query composition to support more powerful searches as describe below.
\end{comment}

\myline{Vector Search Across Multiple Vertex Types} The function supports vector search across multiple vertex types specified in the \texttt{VectorAttributes} argument, as long as the attributes pass the compatibility check. For instance, to find the top-k comments or posts related to a specific topic, you can convert the topic into a vector and execute the following query:

%\begin{minted}[escapeinside=||]{SQL}
\begin{lstlisting}[basicstyle=\fontsize{7.3pt}{8.4pt}\ttfamily]
CREATE QUERY Q1(List<FLOAT> topic_emb, INT k) {
  -- top-k similar messages (comments or posts)
  Msgs = VectorSearch
        ({Comment.content_emb, Post.content_emb}, 
         topic_emb, k);    
  Print Msgs;
}
\end{lstlisting}
%\end{minted}

\myline{Vector Search Function in Query Composition}
Query composition allows the output of one query block to serve as input for another. As discussed earlier, in GSQL, a query procedure consists of a sequence of query blocks, each defined by a \texttt{SELECT-FROM-WHERE} clause. These blocks can produce either a vertex set or a table. When a query block outputs a vertex set, the result can be assigned to a vertex set variable. Subsequent query blocks can reference this variable in their \texttt{FROM} clause, enabling the composition of queries using vertex set variables. 
% \yuxu{enabling the composition of queries using vertex set variables.}%done
% \yuxu{overall I think our use of query composition is a little bit hand waving, not well defined, doesn't read well. but it's Ok for now}

Since the \texttt{VectorSearch} function returns a vertex set, it can participate in query composition. For example, query Q2 below uses the \texttt{VectorSearch} function to retrieve the top-k Comments or Posts closest to a given query.
% For example, query Q2 below uses the \texttt{VectorSearch} function to retrieve the top-k messages (Comments or Posts) closest to a given query embedding. 
% \yuxu{
% For example, query Q2 below uses the \texttt{VectorSearch} function to retrieve the top-k Comments or Posts closest to a given query.
% }%ok
The results are stored in the vertex set variable \texttt{TopKMessages}. In the subsequent query block, this variable is used as the starting point to find the creators of the top-k messages through a 1-hop graph pattern match. 
% \yuxu{As the \texttt{VectorSearch} function returns a vertex set, its output can directly drive the next GSQL query block, achieving efficient query composition.
% }
% \yuxu{the last sentence basically repeats what's stated in the first sentence of this paragraph. Unnecessary, can be removed, or to be merged into the first sentence. Also the previous paragraph already explained how vertex sets (variables) can be used by subsequent query blocks, so this sentence reads rather redundant.}% removed.

%\begin{minted}[escapeinside=||]{SQL}
\begin{lstlisting}[basicstyle=\fontsize{7.3pt}{8.4pt}\ttfamily]
CREATE QUERY Q2(List<FLOAT> topic_emb, INT k) {
  -- top-k similar messages (comments or posts)
  TopKMessages = 
      VectorSearch
          ({Comment.content_emb, Post.content_emb}, 
            topic_emb, k);

  -- find authors via TopKMessages composition
  Authors =
      SELECT p
      FROM (:TopKMessages)-[:hasCreator]->(p:Person);

  PRINT Authors;
}
\end{lstlisting}
%\end{minted}

Conversely, a graph query block can generate a candidate vertex set that is passed to the vector search function as a filter. In the example query Q3 below, the first query block selects comments located in the United States, storing the results in the \texttt{USComments} vertex set. This set is then passed as a filter to the \texttt{VectorSearch} function, achieving query composition.

%\begin{minted}[escapeinside=||]{SQL}
\begin{lstlisting}[basicstyle=\fontsize{7.3pt}{8.4pt}\ttfamily]
CREATE QUERY Q3(List<FLOAT> topic_emb, INT k) {
    -- a global map accumulator
    Map<VERTEX, FLOAT> @@disMap;
    
    -- find U.S. comments from a graph query block
    USComments = 
        SELECT t
        FROM (c:Country)<-[:LOCATED_IN]-(t:Comment)
        WHERE c.name = "United States";

    -- find top-k comments via USComments composition
    TopKComments = 
        VectorSearch
            ({Comment.content_emb}, 
             topic_emb, k, 
             {filter: USComments, ef: 200, 
              distanceMap: @@disMap});

    -- output top-k comments and their distances
    PRINT TopKComments;
    PRINT @@disMap;
}
\end{lstlisting}
%\end{minted}

An interesting variation of Q3 is query Q4, where a Louvain community detection algorithm~\cite{louvain} is first used to tag each person with a community ID. The top-k vector search function is then applied within each community’s vertex set, demonstrating the flexibility of combining vector search with advanced graph analytics.

\begin{lstlisting}[basicstyle=\fontsize{7.3pt}{8.4pt}\ttfamily]
CREATE QUERY Q4(List<FLOAT> topic_emb, INT k) {
    -- detect communities on vertex Person and edge knows
    -- write community id into Person.cid
    C_num = tg_louvain(["Person"], ["knows"]);
    
    -- loop each community to do top-k search 
    FOREACH i IN RANGE[0, C_num] DO
        -- select posts from each community
        CommunityPosts = 
            SELECT t 
            FROM (s:Person)<-[e:hasCreator]-(t:Post)
            WHERE s.cid = i;

        -- do top-k search for each community
        TopKPosts = 
            VectorSearch
              ({Post.content_emb}, 
               topic_emb, k, {filter: CommunityPosts});

        PRINT TopKPosts;
    END;
}
\end{lstlisting}
%\end{minted}
\vspace{-0.1cm}
%\yuxu{-- detect communities on vertex Person and edge knows}
%\yuxu{-- do top-k search for each community}

%\yuxu{The above query gets the number of communities detected as an input parameter (Group\_Num). Better to directly use the number of communities detected by tg\_louvain() inside the query without using the external input parameter.  The number of communities computed by tg\_louvain is in the output json information. To simplify the query, we can assume  x = tg\_louvain() returns the number of communities computed, which should be simple for TG to implement too.}

\myline{Demonstration} Figure~\ref{fig:demo_flat} shows the visual result of the query Q4, in which Person nodes are partitioned into three communities using the Louvain algorithm~\cite{louvain}. For each community, we conduct a top-2 vector search on Posts created by people in this community and present the content of the Posts to analyze the attitudes of different communities toward AI development.
% \yuxu{this Demo section is great!}

% \myline{Final Remarks} Since the GSQL local accumulator can be used to mark any subset of nodes in the graph, we can use a single query block to select all those marked nodes, assign them to a vertex set variable, and feed the variable to the \texttt{VectorSearch} function for vector search. 
% \yuxu{the last sentence looks to readers completely out of place, breaking the flow. We'll only lose points, nothing to gain here. I suggest remove it.}
% Also, we provide a list of built-in functions to support common vector attribute expressions, including \texttt{kth\_element}, \texttt{element\_sum}, \texttt{dimension\_count}, \texttt{norm}, and different distance metric functions. 
% \yuxu{the last sentence's intention is good: showing that we did some more additinal work. However, it's not that interesting, plus we don't have space to explain the meaning of each function.}
% \yuxu{overall, I  suggest that we remove the Final Remarks section, or we have to rewrite it well} % Jianguo: ok, we can remove it

\section{Experiments}\label{sec:experiment}

 % \jianguo{I'll go over this section later}

% \jianguo{If we don't have enough space, this paragraph can be removed}
In this section, we present experimental results to evaluate \sys{}. Our experiments include four parts. The first part is vector search performance evaluation, which aims to test the capability of our system to conduct vector search with high performance and accuracy, compared with the state-of-the-art specialized vector database Milvus~\cite{Milvus21} and two popular graph databases, Neo4j~\cite{Neo4j} and Amazon Neptune~\cite{Neptune}, with the vector search feature (\sect{}~\ref{sec:experiments-vectorsearch}). The second part is scalability evaluation, which aims to test how our system can scale with more machines 
% \yuxu{change nodes to machines as nodes can also mean the nodes in a graph} 
and larger datasets (\sect{}~\ref{sec:experiments-scalability}). The third part is update evaluation, which aims to test the time our system needs to prepare databases from scratch and update existing vectors 
% \yuxu{change databases to vectors} % done
(\sect{}~\ref{sec:experiments-update}). The fourth part is the hybrid vector search and graph search evaluation, in which we prepare a hybrid dataset based on LDBC-SNB (\sect{}~\ref{sec:hybrid_test}).

% \jianguo{To Shige: as we have enough materials, the experimental figures can be made smaller.}

\subsection{Experiment Setting}\label{sec:expsetting}

\myline{Experimental Platform} By default, we perform the experiments using n2d-standard-32 instances (AMD EPYC 7B13, 32 vCPUs, 128GB memory, 64MB L3 cache, SSE) with SCSI disks on Google Cloud.

\myline{Datasets} For the vector search performance evaluation, we use the standard SIFT100M~\cite{SIFTData} and Deep100M~\cite{DEEPData} datasets, each containing 100 million vectors. Additionally, we use SIFT1B~\cite{SIFTData} and Deep1B~\cite{DEEPData} datasets, which contain 1 billion vectors, for scalability evaluation. Table~\ref{tab:dataset} summarizes these datasets.

To evaluate hybrid vector and graph search, we create a new dataset by incorporating vectors into the LDBC SNB benchmark~\cite{ldbc-snb}, as no existing {graph} datasets {with vectors} are available. Specifically, vectors are added as content embeddings to the \texttt{Comment} and \texttt{Post} vertices.  To generate queries, we modify queries from \cite{GSQL}, which simulate a variety of complex read queries for analyzing social network data, and incorporate vector search into them.

% \textcolor{blue}{We choose this synthetic dataset because there is no large-scale enough public hybrid dataset.}

% as there is no such dataset available 
% since we are the first one to do this test, we create a graph and vector hybrid dataset by composing LDBC-SNB and SpaceV.

\begin{table}[tbp]
\small
\centering
\caption{\textbf{Statistics of Datasets}}\label{tab:dataset}
\resizebox{0.95\linewidth}{!}{%
\begin{tabular}{r|r|r|r}\hline\hline
\textbf{Dataset} & \textbf{\# Dimensions} & \textbf{\# Vectors} & \textbf{\# Queries}\\\hline\hline
SIFT100M & 128 & 100,000,000 & 10,000\\\hline
SIFT1B & 128 & 1,000,000,000 & 10,000\\\hline
Deep100M & 96 & 100,000,000  & 10,000 \\\hline
Deep1B & 96 & 1,000,000,000  & 10,000 \\\hline\hline
%Microsoft SpaceV~(10M) & 100 & L2 & 10^{7} & 29,300 \\\hline
%Cohere~(10M) & 768 & IP & 10^{7} & 10,000       
%     \\\hline
\end{tabular}%
}
\vspace{-0.2cm}
\end{table}

\myline{Competitors} To evaluate the vector search performance of \sys{}, we compare it with Neo4j~\cite{Neo4j} and Amazon Neptune~\cite{Neptune}, two popular graph databases that support vector search. Additionally, we compare \sys{} with Milvus~\cite{Milvus21}, a highly optimized 
% \yuxu{and} % Jianguo: should be fine : )
specialized vector database system. For a fair comparison, all systems build the same HNSW index using the same parameters (e.g., $M = 16$ and $efb = 128$ as recommended in \cite{SingleStoreV}).

Since Neo4j and Amazon Neptune do not allow parameter tuning, we present only a single data point in Figure~\ref{fig:recall-QPS} and Figure~\ref{fig:recall-latency}. We use HTTP requests to send vector search queries for all systems except Milvus, as {the performance of} its HTTP port  is extremely slow. Instead, we test Milvus using its gRPC port. For Amazon Neptune, a fully managed cloud service on AWS, we run it using 1024 m-NCU (Neptune Memory Capacity Units)~\cite{NeptuneHardware}, the largest Neptune instance available. This hardware is actually more powerful than the Google Cloud hardware used by \sys{}, ensuring no unfair advantage.

All experiments are conducted in-memory, as a single server has sufficient capacity to store the entire dataset (SIFT100M or Deep100M). For SIFT1B and Deep1B, the datasets are distributed across the memory of multiple servers during the scalability experiment (in \sect{}~\ref{sec:experiments-scalability}).

\subsection{Evaluating Vector Search}\label{sec:experiments-vectorsearch}

%Fig~\ref{fig:recall-QPS-10M} shows that we achieve similar search performance comparing to Milvus on 10M datasets, with QPS ranging from \textbf{80\%} $\sim$ \textbf{114\%}. The lowest recall rate we can adjust is much \textbf{11\%} $\sim$ \textbf{18\%} higher than Neo4j with \textbf{6}$\times$ $\sim$ \textbf{6.9}$\times$ higher QPS. Neo4j has no data in Cohere10M since it doesn't support IP metric. 

% We show the throughput result in Fig~\ref{fig:recall-QPS}, in which we test the throughput of different systems on single machine except Neptune Analytics on 100M datasets. For each test, we forks 16 threads to send requests, and measure the overall time consumption for a system to finish the given query load.

% Fig~\ref{fig:recall-QPS} shows that we achieve \textbf{1.76}$\times$ $\sim$ \textbf{1.06}$\times$ QPS compared to Milvus on 100M datasets thanks to the effective use of multi-core parallelism. We can achieve \textbf{23\%} $\sim$ \textbf{26\%} more recall rate and \textbf{3.77}$\times$ $\sim$ \textbf{5.19}$\times$ higher QPS compared to Neo4j on the lowest recall rate point. As for Neptune Analytics, we apply 1024 m-NCUs for running vector search on 100M datasets. We can achieve \textbf{1.93}$\times$ $\sim$ \textbf{2.7}$\times$ QPS compared to Neptune Analytics, with recall rate slightly lower. Since The information of Neptune Analytics machines is not disclosed, we compare the cost to reach this performance. We are running on one n2d GCP instance with \$1.37 per hour, while 1024 m-NCUs cost \$30.72 per hour, which is \textbf{22.42}$\times$ more expense. 

\begin{figure}[tbp]
\centering
\includegraphics[width=0.45\textwidth]{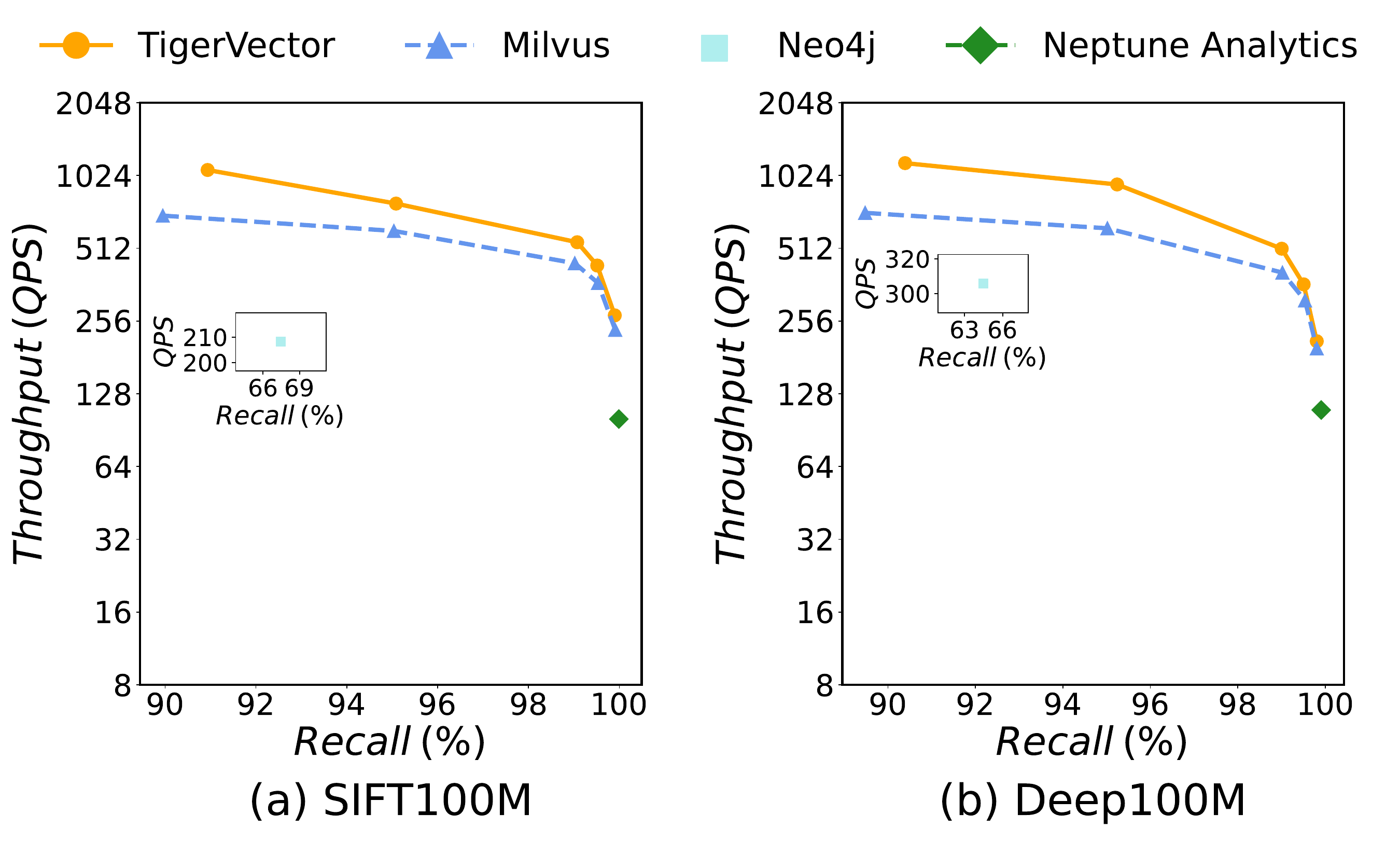}
\caption{Throughput Evaluation on SIFT100M and Deep100M}
\label{fig:recall-QPS}
\vspace{-0.1cm}
\end{figure}

Figure~\ref{fig:recall-QPS} shows the throughput results on SIFT100M and Deep100M, using 16 threads to send queries. The results demonstrate that \sys{} significantly outperforms both Neo4j and Amazon Neptune in vector search performance. 
% \yuxu{"for vector search" ==> in vector search performance} % done
Specifically, on SIFT100M, \sys{} simultaneously  delivers \textbf{5.19}$\times$ higher QPS and \textbf{23\%} higher recall \textbf{(90.94\%, 1079.00)} compared to Neo4j \textbf{(67.50\%, 208.15)}. On Deep100M, \sys{} achieves \textbf{3.77}$\times$ higher QPS and \textbf{26\%} higher recall \textbf{(90.39\%, 1151.92)} compared to Neo4j \textbf{(64.46\%, 305.77)}.
Moreover, \sys{} achieves \textbf{1.93}$\times$ to \textbf{2.7}$\times$ higher throughput compared to Amazon Neptune while maintaining a similar recall rate (\textbf{99.9\%}), even though Amazon Neptune uses better hardware. From a cost perspective, \sys{} runs on an n2d GCP instance costing \$1.37 per hour, whereas Amazon Neptune uses 1024 m-NCUs costing \$30.72 per hour, which is \textbf{22.42}$\times$ more expensive.

Interestingly, when compared to Milvus, a highly optimized 
% \yuxu{and} % Jianguo: should be fine : )
specialized vector database, \sys{} remains competitive, achieving \textbf{1.07}$\times$ to \textbf{1.61}$\times$ higher throughput. This can be attributed to the more effective use of multi-core parallelism, as TigerGraph is an MPP database. Another factor is the difference in programming languages, as Milvus is written in Go, whereas \sys{} is written in C++.

\begin{figure}[tbp]
\centering
\includegraphics[width=0.45\textwidth]{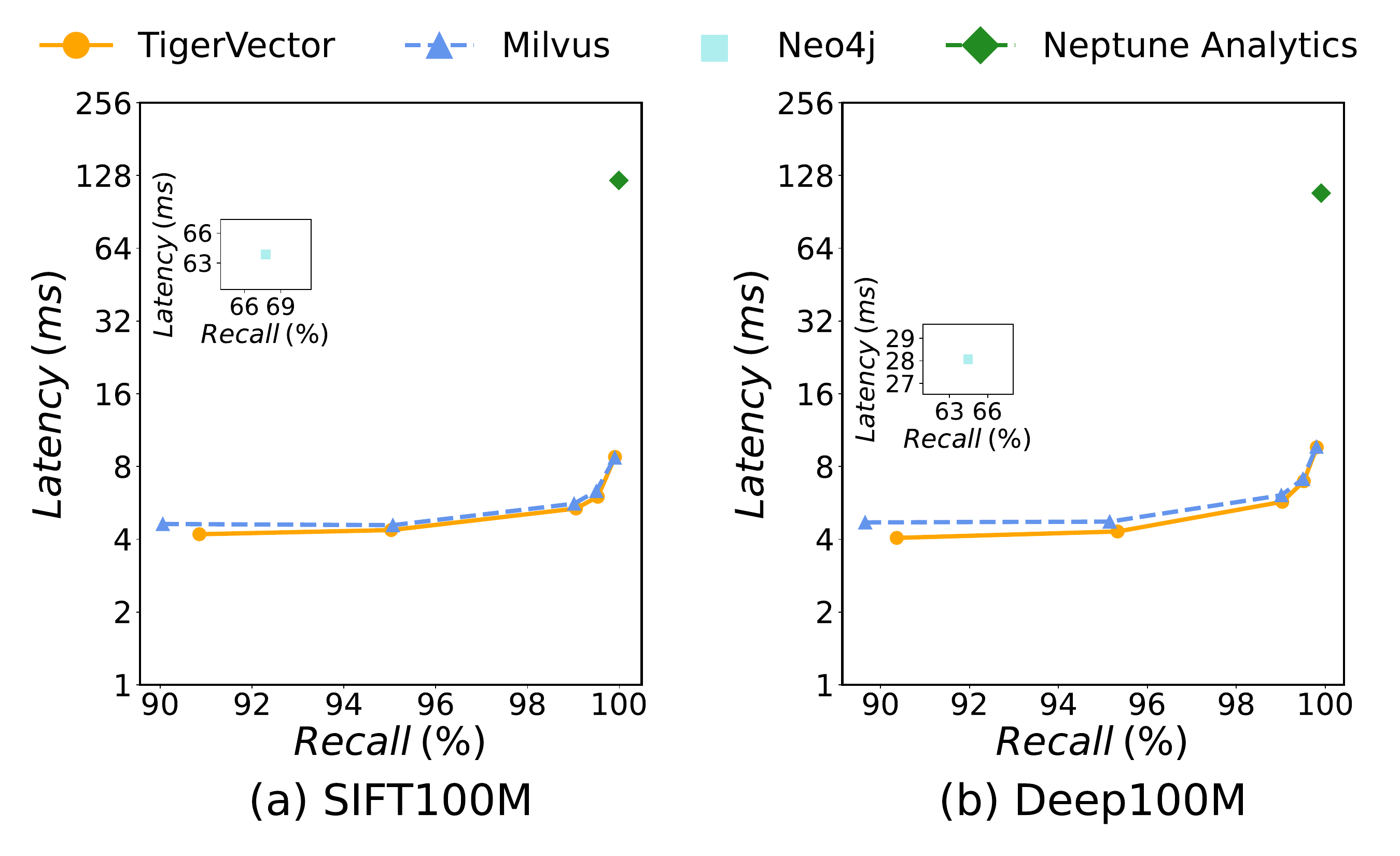}
\caption{Latency Evaluation on SIFT100M and Deep100M}
\label{fig:recall-latency}
\vspace{-0.2cm}
\end{figure}

Figure~\ref{fig:recall-latency} shows the latency results obtained with a single thread. We observe a trend similar to that in Figure~\ref{fig:recall-QPS}. 
% \yuxu{We observe a trend similar to that in Figure ~\ref{fig:recall-QPS}.} %done
The results show that \sys{} significantly outperforms Neo4j and Amazon Neptune, being up to \textbf{15}$\times$ and \textbf{13.9}$\times$ faster, respectively. Furthermore, \sys{} is slightly faster than Milvus, achieving up to \textbf{1.16}$\times$ lower latency.

\begin{comment}
%As many customers care about the predictable memory consumption, Figure~\ref{fig:long_term_usage} shows the memory utilization over five epochs, where each epoch processes a batch of different query vectors. The results demonstrate that \sys{} maintains predictable memory usage. 

\begin{figure}[tbp]
\centering
\includegraphics[width=0.5\textwidth]{figures/chart/long_term_usage.pdf}
\caption{Memory Usage Under Pressure}
\label{fig:long_term_usage}
% \vspace{-0.3cm}
\end{figure}

\end{comment}

\begin{comment}

\begin{figure}[tbp]
\centering
\includegraphics[width=0.5\textwidth]{figures/chart/100M-recall-latency.png}
\caption{Latency Test on 100M Datasets}
\label{fig:100M-recall-latency}
% \vspace{-0.3cm}
\end{figure}
\end{comment}

\subsection{Evaluating Scalability}\label{sec:experiments-scalability}

In this experiment, we evaluate the scalability of \sys{} in terms of the number of nodes and data size. To minimize the impact of the network port, we use an additional sender machine that evenly distributes requests across all machines. The sender machine utilizes a popular HTTP benchmarking tool, wrk2~\cite{wrk2}. In this benchmark, it maintains 320 connections and forks 16 threads in total. Each thread prepares payloads with randomly selected query vectors to minimize the impact of CPU caches caused by identical payloads. The number of requests sent per second is sufficiently high to saturate the throughput of our system.

\begin{figure}[tbp]
\centering
\includegraphics[width=0.45\textwidth]{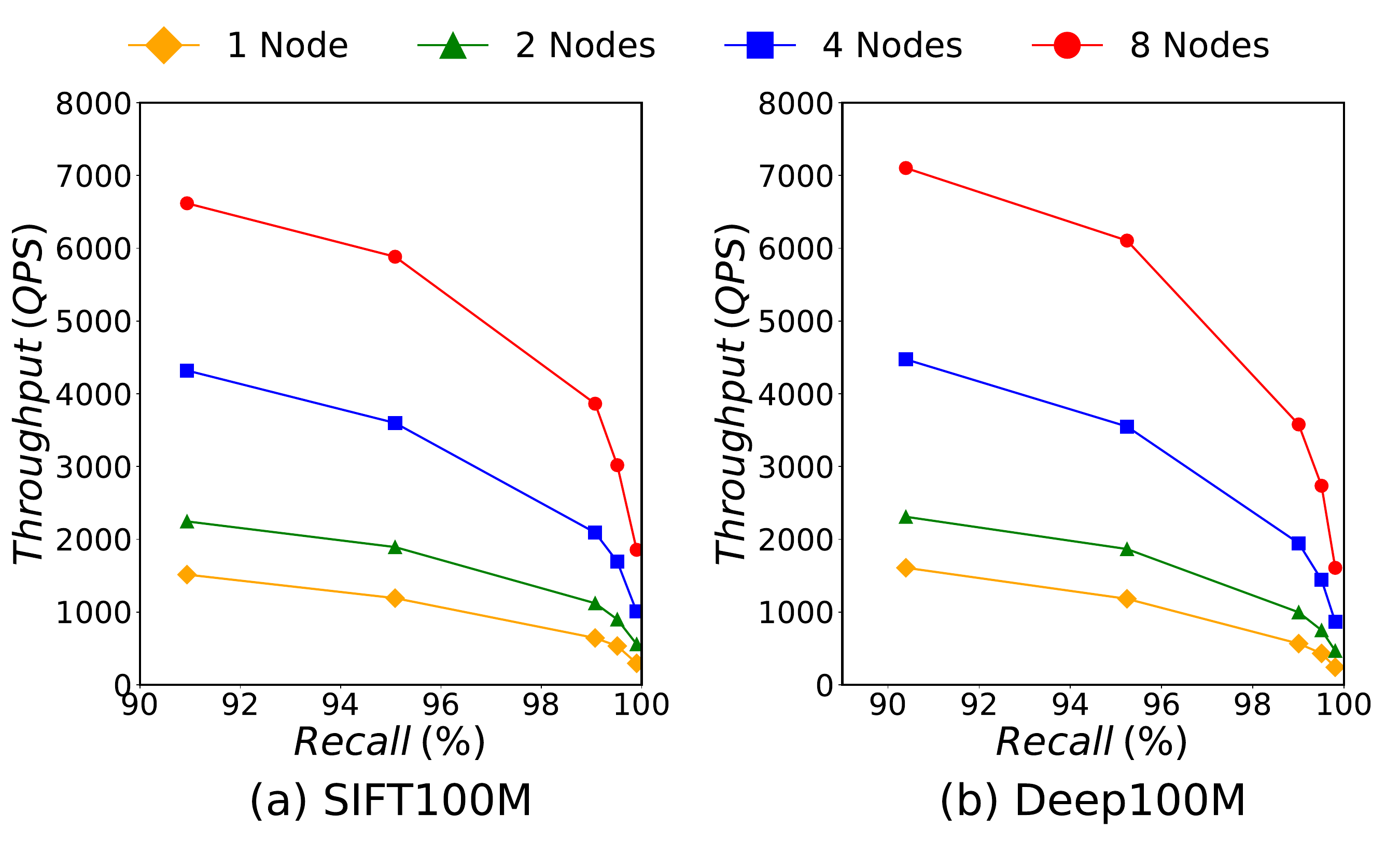}
\caption{Node Scalability}
\label{fig:node_scalability}
\vspace{-0.1cm}
\end{figure}

Figure~\ref{fig:node_scalability} illustrates the node scalability of \sys{}. The results show that \sys{} achieves excellent scalability. For instance, when the recall rate is 99.9\%, \sys{} achieves a throughput performance gain of \textbf{1.84}$\times$ to \textbf{1.91}$\times$ when the number of machines is doubled. Even with a recall rate of 90\%, where network communication overhead constitutes a larger proportion compared to the high recall rate scenario, it achieves a \textbf{1.5}$\times$ performance gain when the number of machines is doubled.

\begin{figure}[tbp]
\centering
\includegraphics[width=0.4\textwidth]{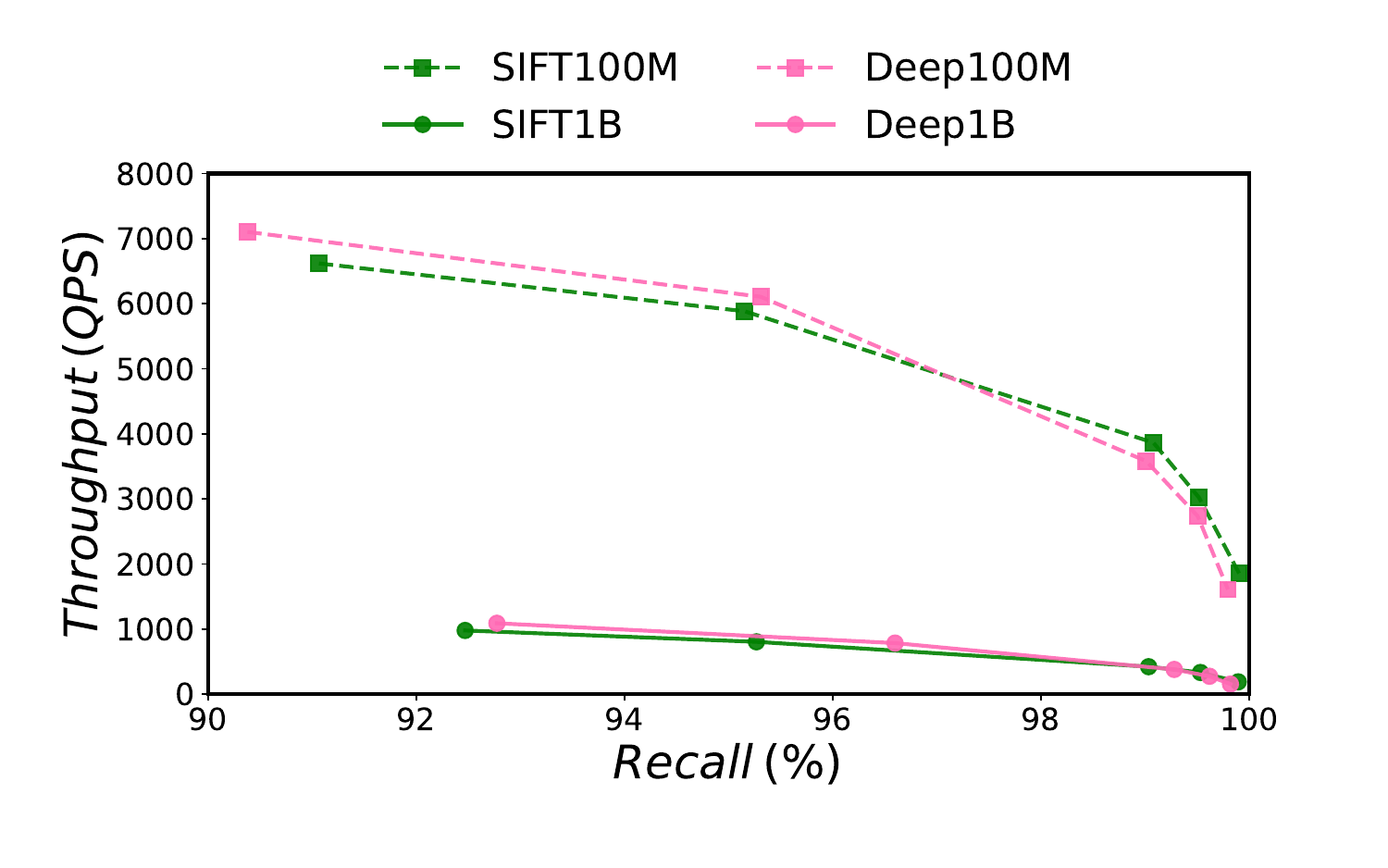}
\caption{Data Size Scalability}
\label{fig:datasize_scalability}
\vspace{0.25cm}
\end{figure}

Figure~\ref{fig:datasize_scalability} shows the scalability of \sys{} with respect to dataset size, where we scale the dataset size from 100M to 1B while keeping the same search parameters. We use 8 machines to run this experiment and utilize wrk2 for benchmarking throughput, maintaining the same configuration as in Figure~\ref{fig:node_scalability}.

The results in Figure~\ref{fig:datasize_scalability} show that \sys{} achieves great scalability, with throughput decreasing roughly proportionally as the dataset size increases. The number of segments for the 1B dataset is exactly 10$\times$ that of the 100M dataset. For performance with $ef = 12$ (the lowest recall rate point), when the dataset scales from SIFT100M to SIFT1B, \sys{} still maintains \textbf{14.75\%} of the original QPS despite a 10$\times$ increase in workload. This is because the proportion of computation increases, improving CPU utilization from 60\% to 80\%. 
%The recall rate also improves \shige{under the same $ef$} because more segments contribute their local search results.
%\yuxu{the recall rate improves when ??? isn't changed? I don't understand the last sentence}\shige{When using the same ef, scale from 100M to 1B, the QPS goes down and recall rates goes up.} 
%\yuxu { I see, it's not clear from looking at Fig 10. does it make sense to add a little more info to say something like "The recall rate also improves from X to Y usign ef = Z? so that readers won't need to think hard about this? }
%\shige{maybe we can just delete this sentence, as I recheck the numbers, just increase recall rate very slightly.}
%\yuxu{removing it is good, it doesn't help much and could raise questions}
For higher recall rate points, the QPS decreases proportionally to \textbf{10\%} when the dataset size increases by \textbf{10}$\times$.

\subsection{Evaluating Index Update}\label{sec:experiments-update}

\begin{table}[tbp]
\small
\centering
\caption{\textbf{Index Building Time}}\label{tab:index_building}
\resizebox{0.95\linewidth}{!}{%
\begin{tabular}{c|c|c|c|c}\hline\hline
\textbf{Dataset} &\textbf{Measure}& \textbf{\sys{}} & \textbf{Milvus} & \textbf{Neo4j}\\\hline\hline
         & End to End & \textbf{3,977s} & 8,577s & 20,679s\\
SIFT100M  & Data Load & 202s & 4,554s & 133s \\
        & Index Build & 3,775s & 4,023s & 20,546s \\\hline
         
         & End to End & \textbf{3,462s} & 6,430s  & 23,560s \\
Deep100M  & Data Load& 338s & 3,258s  & 559s \\
         & Index Build  & 3,124s & 3,172s  & 23,001s \\\hline\hline
         
%Microsoft SpaceV~(10M) & 100 & L2 & 10^{7} & 29,300 \\\hline
%Cohere~(10M) & 768 & IP & 10^{7} & 10,000       
%     \\\hline

\end{tabular}%
}
\end{table}

\begin{figure}[tbp]
\centering
\includegraphics[width=0.4\textwidth]{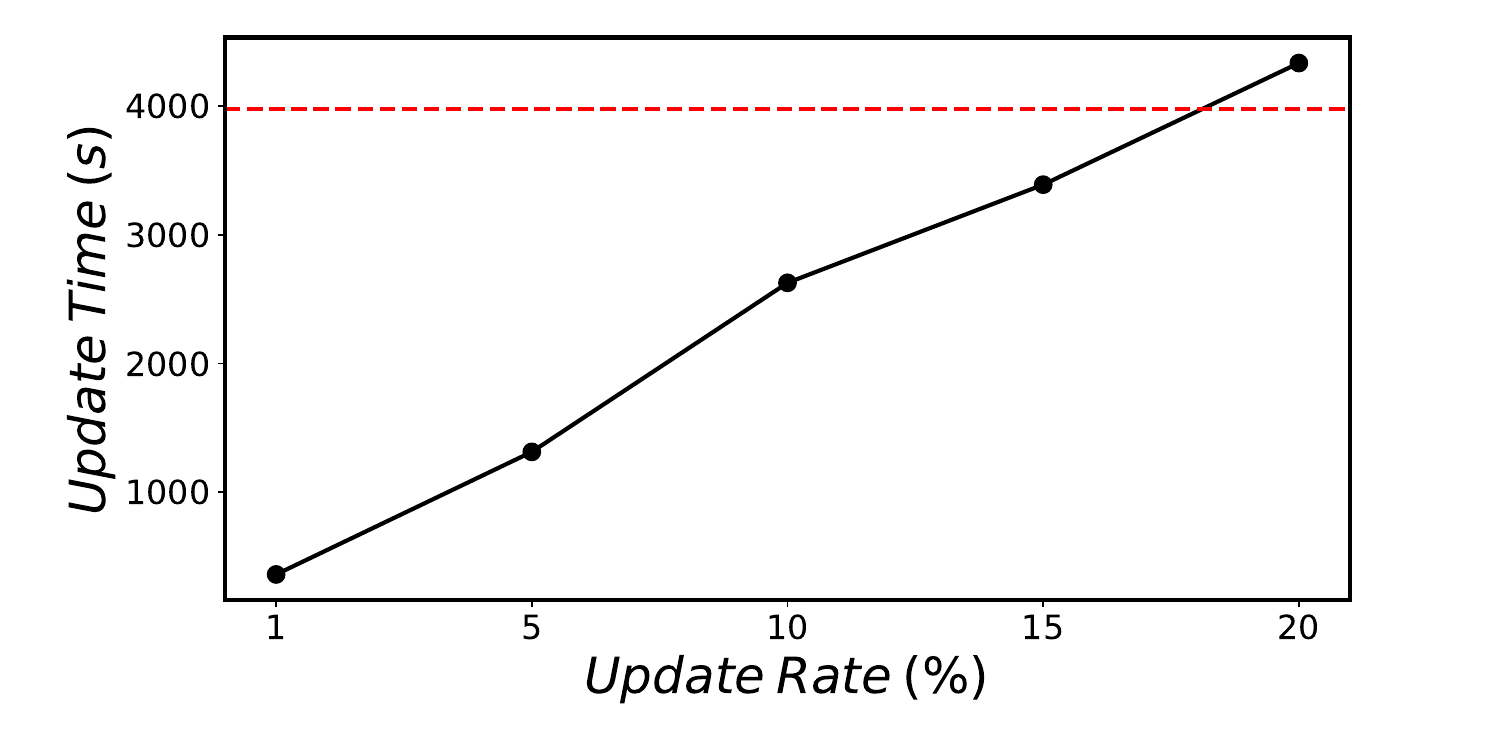}
\caption{Index Update Evaluation on SIFT100M}
\label{fig:update_insert}
%\vspace{0.2cm}
\end{figure}

In this experiment, we evaluate the vector index building and update performance of \sys{}. We compare the end-to-end index building time across different datasets using a single machine. \sys{} and Neo4j both load data from CSV files using their respective loading tools, while Milvus loads data directly from raw vector files. The HNSW vector index is built after data loading.

Table~\ref{tab:index_building} shows that \sys{} achieves \textbf{5.2}$\times$ to \textbf{6.8}$\times$ shorter index building time compared to Neo4j. Additionally, \sys{} is \textbf{1.86}$\times$ to \textbf{2.16}$\times$ faster than Milvus for index building. This improvement is due to \sys{}'s optimized data loading tool, which outperforms Milvus. Although Neo4j has a similar data loading time as \sys{}, it exhibits slower index building performance.

Figure~\ref{fig:update_insert} evaluates the incremental index update performance of \sys{}. It shows that as the update ratio increases, the update time also increases. The red line indicates the time required to completely rebuild the HNSW index in \sys{}. The results suggest that if more than 20\% of the vectors are updated, it is more efficient to rebuild the index rather than updating it incrementally.

% We compares the end to end index building time on different datasets on 100M datasets using one machine. \sys{} and Neo4j both load data from csv files using their loading tool, and Milvus loads data directly from raw vector files. HNSW vector indexes are built after data loading. 

% Table~\ref{tab:index_building} shows that \sys{} can have \textbf{1.86}$\times$ $\sim$ \textbf{2.16}$\times$ faster building time compared to Milvus, and is \textbf{5.2}$\times$ $\sim$ \textbf{6.8}$\times$ faster than Neo4j index building. We can observe from the Index Build line that, \sys{} and Milvus actually has similar HNSW index building time, but \sys{} gains faster loading speed due to optimized loading tool. Neo4j has similar loading time compared to \sys{}, but it has poor index building performance.

% Fig~\ref{fig:update_insert} compares the end to end vector update time under different update rate using SIFT100M. We can observe that the update time scales linear with update rete, because each vector record update requires a top-k search within a segment. The red horizontal line marks the time consumption for totally rebuild the whole dataset, and it tells that we would better choose to rebuild if we need to update 20\% or more vectors at once.

%\begin{figure}[tbp]
%\centering
%\includegraphics[width=0.5\textwidth]{figures/chart/off_preparation_time.pdf}
%\caption{Index Building Evaluation}
%\label{fig:index_building_time}
% \vspace{-0.3cm}
%\end{figure}

\subsection{Evaluating Hybrid Search}\label{sec:hybrid_test}

\begin{comment}

Table~\ref{tab:performance_metrics_sf10} and Table~\ref{tab:performance_metrics_sf30} present the hybrid search performance of \sys{}. We use a hybrid dataset based on LDBC-SNB and add vectors to the Post and Comment nodes as content vectors. The workload is derived from LDBC-SNB (i.e., IC (Interactive Complex) queries) and modified to include vector search on the Comment and Post nodes traversed during the queries. We measure the execution time of the queries under different dataset sizes and different numbers of hops in the IC queries. The number of Messages (Comments and Posts) traversed for vector search varies across different queries. For instance, IC5 traverses the most Messages because it reaches more nodes during traversal. IC6 and IC11 traverse a moderate number of nodes, while IC3 traverses the fewest. IC9 performs vector search on only 20 Messages, as it sets a limit of 20 for graph queries.

From the table results, we observe that as the number of hops in graph queries or the dataset size increases, the end-to-end execution time (comprising graph query execution time and vector search time) also increases. The number of Messages traversed by a query (#candidate) grows as well. However, the vector search time for each query remains limited to a few milliseconds, which is negligible compared to the overall query execution time.
\end{comment}

%\mingxi{version todo: find root cause on why IC11 is slower than IC5}
In this experiment, we evaluate the performance of hybrid queries by modifying the interactive complex (IC) family of queries from the LDBC SNB benchmark~\cite{ldbc-snb} to incorporate top-k vector search. Specifically, we select IC queries involving the \texttt{KNOWS} edge type and vary the number of repetitions of \texttt{KNOWS} to generate queries with different path lengths. A global accumulator collects all the \texttt{Message} vertices (either \texttt{Post} or \texttt{Comment}) matched by the IC queries. Subsequently, a top-k vector search is performed on the collected \texttt{Message} set. This design ensures that each graph-shaped IC query touches a different number of segments, allowing us to study the behavior of our top-k index.

The benchmark is conducted at scale factors of 10 and 30, 
% \yuxu{readers may not be familiar with the factors, need to add the sizes of the datasets} 
with each \texttt{Message} vertex augmented by an embedding attribute sampled from the SIFT100M dataset. Tables~\ref{tab:performance_metrics_sf10} and~\ref{tab:performance_metrics_sf30} summarize the hybrid search performance of \sys{}.

We measure the end-to-end query execution times, the number of collected \texttt{Message} candidates, and the top-k vector search time. The queries have different query shapes and path lengths (hops), which result in varying sizes of the \texttt{Message} set used for the top-k vector search. For example, IC5 collects the largest number (5M to 17M) of \texttt{Messages} due to its broader node traversal, whereas IC6 and IC11 collect a moderate number of nodes, and IC3 and IC9 operate on fewer than 100 \texttt{Messages}.
%\yuxu{ not sure if bind/binds is the right word.  typically we say bind to (variables?).   What about we use collect? We used collect earlier}

The results show that the end-to-end execution time increases with the number of hops, typically scaling 
% \yuxu{either} 
linearly or sublinearly. Notably, vector search demonstrates excellent performance and scalability, often completing within a few milliseconds. Interestingly, the proportion of time spent on top-k vector search does not always scale directly with the size of the candidate \texttt{Message} set. For instance, while 3-hop-IC5 has a larger candidate set than 3-hop-IC11 on SF10 (6,598,427 vs. 49,475), it spends less time on vector search (1.52ms vs. 2.32ms). We find that IC5 touched 32 segments while IC11 touched 64 segments. We also note that even though IC11 performed brute-force search on its segments while IC5 invoked HNSW index search, the time spent on each segment is similar, which explains the shorter vector search time for IC5.

\begin{table}[tbp]
\small
\centering
\caption{Evaluating Hybrid Search for SF10}\label{tab:performance_metrics_sf10}
\scalebox{0.85}{
\begin{tabular}{c|c|c|c|c|c|c}
\hline\hline
\textbf{\#Hops} & \textbf{Measure} & \textbf{IC3} & \textbf{IC5} & \textbf{IC6} & \textbf{IC9} & \textbf{IC11} \\ \hline\hline
                           & End to End  &0.64s	&2.36s	&1.22s	&0.88s	&0.67s         \\
                  2         & \#candidate   &0	     &5,266,252 &1,261	&20	     &11,242           \\
                          & Vector Search & 0          & \textbf{1.18ms }     &\textbf{0.31ms}         &\textbf{0.15ms}         & \textbf{0.85ms}         \\ \hline
                           & End to End  &1.15s	&3.19s	&2.44s	&2.62s	&0.91s        \\
                  3         & \#candidate   &15	&6,598,427	&2,724	&20	&49,475       \\
                           & Vector Search & \textbf{0.13ms}         &\textbf{1.52ms}    &\textbf{0.45ms}         &\textbf{0.16ms}         &\textbf{2.32ms}           \\ \hline
                               & End to End  &1.56s	&3.37s	&2.66s	&2.88s	&1.13s         \\
                  4         & \#candidate    &23	&6,601,992	&2,726	&20	&50,036         \\
                           & Vector Search & \textbf{0.15ms}         &\textbf{1.60ms}         &\textbf{0.40ms}        &\textbf{0.17ms}         &\textbf{2.83ms}          \\ \hline\hline
\end{tabular}
}
\end{table}
\vspace{-0.1cm}

\begin{table}[tbp]
\small
\centering
\caption{Evaluating Hybrid Search for SF30}\label{tab:performance_metrics_sf30}
\scalebox{0.85}{
\begin{tabular}{c|c|c|c|c|c|c}
\hline\hline
 \textbf{\#Hops} & \textbf{Measure} & \textbf{IC3} & \textbf{IC5} & \textbf{IC6} & \textbf{IC9} & \textbf{IC11} \\ \hline\hline
                           & End to End  &1.10s	&4.78s	&1.44s	&1.83s	&0.78s         \\
               2            & \#candidate    &3	&11,679,980	&1,603	&20	&21,519           \\
                          & Vector Search & \textbf{0.09ms}       & \textbf{2.32ms}      & \textbf{0.36ms}  &\textbf{0.23ms}        & \textbf{1.32ms}         \\ \hline
                              & End to End  &3.30s	&10.67s	&8.31s	&6.94s	&1.59s        \\
                3           & \#candidate    &71	&17,543,581	&8,348	&20	&118,893       \\
                           & Vector Search & \textbf{0.20ms}         &\textbf{2.23ms}         &\textbf{0.96ms}        &\textbf{0.17ms}        &\textbf{5.44ms}           \\ \hline
                           & End to End  & 3.65s	&11.26s	&9.00s	&7.78s	&2.12s        \\
                 4          & \#candidate    & 71	&17,567,224	&8,364	&20	&122,459        \\
                           & Vector Search & \textbf{0.21ms}        &\textbf{3.92ms}        &\textbf{0.86ms}         &\textbf{0.22ms}         &\textbf{4.94 ms}         \\ \hline\hline
\end{tabular}
}
\vspace{-0.1cm}
\end{table}

% \newpage

\section{Lessons and Discussion}\label{sec:discussion}

In this section, we discuss the key lessons learned from integrating vector search within TigerGraph.
% which can also be applied to other graph databases, such as Neo4j and Amazon Neptune.

\vspace{-0.1cm}
\myline{Decoupling vector embeddings from other graph attributes} There are several advantages to separating vector embeddings from other graph attributes: (1) We can leverage fast, native vector index implementations to support vector search, achieving performance comparable to highly optimized 
% \yuxu{and} % Jianguo: should be fine : )
specialized vector databases. (2) We can conveniently manage different types of vector embeddings using the \texttt{embedding} type. (3) We can support efficient vector updates without negatively impacting existing graph database updates. (4) We can reduce vector data movement during updates to other attributes in graph databases.

\vspace{-0.1cm}
\myline{MPP architecture} 
% \yuxu{ just MPP (without native)}
Many graph databases already support parallel or distributed graph processing via MPP. When integrating vector search into such graph databases, we can leverage MPP to support parallel and distributed vector search by building an index for each segment.

\vspace{-0.1cm}
\myline{Query composition} Query composition is a powerful tool for expressing complex queries. It seamlessly integrates declarative vector search, the vector search API, and other graph algorithms to support advanced RAGs.

\vspace{-0.1cm}
\myline{Unified design} We show that it is possible to support vector search within a graph database while achieving high performance. Therefore, we advocate for a unified system that simultaneously supports both vector search and graph search. Such a unified design offers additional advantages, including enabling hybrid vector and graph searches for advanced RAGs and addressing issues related to consistency, data silos, and data movement.

\vspace{-0.1cm}
\myline{Extension to other graph databases} We believe that the above lessons are applicable not only to TigerGraph but also to other graph databases, such as Neo4j and Amazon Neptune.

\vspace{-0.12cm}
\section{Conclusion}\label{sec:conclusion}

In this work, we present \sys{}, a new system that seamlessly integrates vector search into TigerGraph. By unifying the management of graph and vector data, \sys{} can support advanced RAGs and can serve as a data retrieval infrastructure for LLMs. \sys{} introduces a suite of techniques to address challenges in performance and usability. 
Extensive experiments show that \sys{} achieves performance on par with or better than Milvus and significantly outperforms Neo4j and Amazon Neptune in vector search.

\section{Acknowledgments}
Jianguo Wang acknowledges the support of the National Science Foundation under Grant Number \href{https://www.nsf.gov/awardsearch/showAward?AWD_ID=2337806}{2337806}.

% \clearpage
% \newpage
% \balance
\bibliographystyle{ACM-Reference-Format}
\bibliography{paper}


\begin{thebibliography}{39}

%%% the Web address.

\ifx \showCODEN    \undefined \def \showCODEN     #1{\unskip}     \fi
\ifx \showDOI      \undefined \def \showDOI       #1{#1}\fi
\ifx \showISBNx    \undefined \def \showISBNx     #1{\unskip}     \fi
\ifx \showISBNxiii \undefined \def \showISBNxiii  #1{\unskip}     \fi
\ifx \showISSN     \undefined \def \showISSN      #1{\unskip}     \fi
\ifx \showLCCN     \undefined \def \showLCCN      #1{\unskip}     \fi
\ifx \shownote     \undefined \def \shownote      #1{#1}          \fi
\ifx \showarticletitle \undefined \def \showarticletitle #1{#1}   \fi
\ifx \showURL      \undefined \def \showURL       {\relax}        \fi
% invisible to TeX
\providecommand\bibfield[2]{#2}
\providecommand\bibinfo[2]{#2}
\providecommand\natexlab[1]{#1}
\providecommand\showeprint[2][]{arXiv:#2}

\bibitem[LDB({[n.\,d.]})]%
        {LDBC}
 \bibinfo{year}{[n.\,d.]}\natexlab{}.
\newblock \bibinfo{title}{{LDBC (\url{https://ldbcouncil.org/benchmarks/overview/})}}.
\newblock
\newblock


\bibitem[SIF({[n.\,d.]})]%
        {SIFTData}
 \bibinfo{year}{[n.\,d.]}\natexlab{}.
\newblock \bibinfo{title}{{SIFTData (\url{http://corpus-texmex.irisa.fr/})}}.
\newblock
\newblock


\bibitem[Neo({[n.\,d.]})]%
        {Neo4jVecLuceneIndex}
 \bibinfo{year}{[n.\,d.]}\natexlab{}.
\newblock \bibinfo{title}{{Vector Indexes of Neo4j (\url{https://neo4j.com/docs/cypher-manual/current/indexes/semantic-indexes/vector-indexes/})}}.
\newblock
\newblock


\bibitem[Wea({[n.\,d.]})]%
        {Weaviate}
 \bibinfo{year}{[n.\,d.]}\natexlab{}.
\newblock \bibinfo{title}{{Weaviate: An Open Source Vector Database (\url{https://github.com/weaviate/weaviate})}}.
\newblock
\newblock


\bibitem[wrk({[n.\,d.]})]%
        {wrk2}
 \bibinfo{year}{[n.\,d.]}\natexlab{}.
\newblock \bibinfo{title}{{wrk2 (\url{https://github.com/giltene/wrk2})}}.
\newblock
\newblock


\bibitem[Nep(2024)]%
        {NeptuneHardware}
 \bibinfo{year}{2024}\natexlab{}.
\newblock \bibinfo{title}{{Amazon Neptune Analytics now introduces new smaller capacity units (\url{https://aws.amazon.com/about-aws/whats-new/2024/07/amazon-neptune-analytics-smaller-capacity-units/})}}.
\newblock
\newblock


\bibitem[Angles et~al\mbox{.}(2023)]%
        {AnglesBD0GHLLMM23}
\bibfield{author}{\bibinfo{person}{Renzo Angles}, \bibinfo{person}{Angela Bonifati}, \bibinfo{person}{Stefania Dumbrava}, \bibinfo{person}{George Fletcher}, \bibinfo{person}{Alastair Green}, \bibinfo{person}{Jan Hidders}, \bibinfo{person}{Bei Li}, \bibinfo{person}{Leonid Libkin}, \bibinfo{person}{Victor Marsault}, \bibinfo{person}{Wim Martens}, \bibinfo{person}{Filip Murlak}, \bibinfo{person}{Stefan Plantikow}, \bibinfo{person}{Ognjen Savkovic}, \bibinfo{person}{Michael Schmidt}, \bibinfo{person}{Juan Sequeda}, \bibinfo{person}{Slawek Staworko}, \bibinfo{person}{Dominik Tomaszuk}, \bibinfo{person}{Hannes Voigt}, \bibinfo{person}{Domagoj Vrgoc}, \bibinfo{person}{Mingxi Wu}, {and} \bibinfo{person}{Dusan Zivkovic}.} \bibinfo{year}{2023}\natexlab{}.
\newblock \showarticletitle{{PG-Schema: Schemas for Property Graphs}}.
\newblock \bibinfo{journal}{\emph{Proceedings of the ACM on Management of Data (PACMMOD)}} \bibinfo{volume}{1}, \bibinfo{number}{2} (\bibinfo{year}{2023}), \bibinfo{pages}{198:1--198:25}.
\newblock


\bibitem[Babenko and Lempitsky(2016)]%
        {DEEPData}
\bibfield{author}{\bibinfo{person}{Artem Babenko} {and} \bibinfo{person}{Victor~S. Lempitsky}.} \bibinfo{year}{2016}\natexlab{}.
\newblock \showarticletitle{{Efficient Indexing of Billion-Scale Datasets of Deep Descriptors}}. In \bibinfo{booktitle}{\emph{{IEEE} Conference on Computer Vision and Pattern Recognition (CVPR)}}. \bibinfo{pages}{2055--2063}.
\newblock


\bibitem[Blondel et~al\mbox{.}(2008)]%
        {louvain}
\bibfield{author}{\bibinfo{person}{Vincent~D Blondel}, \bibinfo{person}{Jean-Loup Guillaume}, \bibinfo{person}{Renaud Lambiotte}, {and} \bibinfo{person}{Etienne Lefebvre}.} \bibinfo{year}{2008}\natexlab{}.
\newblock \showarticletitle{{Fast Unfolding of Communities in Large Networks}}.
\newblock \bibinfo{journal}{\emph{Journal of Statistical Mechanics: Theory and Experiment}} \bibinfo{volume}{2008}, \bibinfo{number}{10} (\bibinfo{year}{2008}), \bibinfo{pages}{P10008}.
\newblock


\bibitem[Chen et~al\mbox{.}(2024)]%
        {SingleStoreV}
\bibfield{author}{\bibinfo{person}{Cheng Chen}, \bibinfo{person}{Chenzhe Jin}, \bibinfo{person}{Yunan Zhang}, \bibinfo{person}{Sasha Podolsky}, \bibinfo{person}{Chun Wu}, \bibinfo{person}{Szu{-}Po Wang}, \bibinfo{person}{Eric Hanson}, \bibinfo{person}{Zhou Sun}, \bibinfo{person}{Robert Walzer}, {and} \bibinfo{person}{Jianguo Wang}.} \bibinfo{year}{2024}\natexlab{}.
\newblock \showarticletitle{{SingleStore-V: An Integrated Vector Database System in SingleStore}}.
\newblock \bibinfo{journal}{\emph{Proceedings of the VLDB Endowment (PVLDB)}} \bibinfo{volume}{17}, \bibinfo{number}{12} (\bibinfo{year}{2024}), \bibinfo{pages}{3772--3785}.
\newblock


\bibitem[Derbier(2023)]%
        {DgraphVector}
\bibfield{author}{\bibinfo{person}{Raphael Derbier}.} \bibinfo{year}{2023}\natexlab{}.
\newblock \bibinfo{title}{{Dgraph and Vector Database - The Best of Two Worlds (\url{https://dgraph.io/blog/post/20230628-vectordb/})}}.
\newblock
\newblock


\bibitem[Deutsch et~al\mbox{.}(2019)]%
        {tigergraph}
\bibfield{author}{\bibinfo{person}{Alin Deutsch}, \bibinfo{person}{Yu Xu}, \bibinfo{person}{Mingxi Wu}, {and} \bibinfo{person}{Victor~E. Lee}.} \bibinfo{year}{2019}\natexlab{}.
\newblock \showarticletitle{{TigerGraph: A Native {MPP} Graph Database}}.
\newblock \bibinfo{journal}{\emph{CoRR}}  \bibinfo{volume}{abs/1901.08248} (\bibinfo{year}{2019}).
\newblock


\bibitem[Deutsch et~al\mbox{.}(2020)]%
        {GSQL}
\bibfield{author}{\bibinfo{person}{Alin Deutsch}, \bibinfo{person}{Yu Xu}, \bibinfo{person}{Mingxi Wu}, {and} \bibinfo{person}{Victor~E. Lee}.} \bibinfo{year}{2020}\natexlab{}.
\newblock \showarticletitle{Aggregation Support for Modern Graph Analytics in TigerGraph}. In \bibinfo{booktitle}{\emph{Proceedings of the ACM International Conference on Management of Data (SIGMOD)}}. \bibinfo{pages}{377--392}.
\newblock


\bibitem[Edge et~al\mbox{.}(2024)]%
        {edge2024local}
\bibfield{author}{\bibinfo{person}{Darren Edge}, \bibinfo{person}{Ha Trinh}, \bibinfo{person}{Newman Cheng}, \bibinfo{person}{Joshua Bradley}, \bibinfo{person}{Alex Chao}, \bibinfo{person}{Apurva Mody}, \bibinfo{person}{Steven Truitt}, {and} \bibinfo{person}{Jonathan Larson}.} \bibinfo{year}{2024}\natexlab{}.
\newblock \showarticletitle{{From Local to Global: {A} Graph {RAG} Approach to Query-Focused Summarization}}.
\newblock \bibinfo{journal}{\emph{CoRR}}  \bibinfo{volume}{abs/2404.16130} (\bibinfo{year}{2024}).
\newblock


\bibitem[Eifrem(2024)]%
        {Neo4jRAG2}
\bibfield{author}{\bibinfo{person}{Emil Eifrem}.} \bibinfo{year}{2024}\natexlab{}.
\newblock \bibinfo{title}{{GraphRAG: The Marriage of Knowledge Graphs and RAG (\url{https://www.youtube.com/watch?v=knDDGYHnnSI})}}.
\newblock
\newblock


\bibitem[Fan et~al\mbox{.}(2024)]%
        {FanDNWLYCL24}
\bibfield{author}{\bibinfo{person}{Wenqi Fan}, \bibinfo{person}{Yujuan Ding}, \bibinfo{person}{Liangbo Ning}, \bibinfo{person}{Shijie Wang}, \bibinfo{person}{Hengyun Li}, \bibinfo{person}{Dawei Yin}, \bibinfo{person}{Tat{-}Seng Chua}, {and} \bibinfo{person}{Qing Li}.} \bibinfo{year}{2024}\natexlab{}.
\newblock \showarticletitle{{A Survey on {RAG} Meeting LLMs: Towards Retrieval-Augmented Large Language Models}}. In \bibinfo{booktitle}{\emph{Proceedings of the ACM Conference on Knowledge Discovery and Data Mining (KDD)}}. \bibinfo{pages}{6491--6501}.
\newblock


\bibitem[Gartner5188263(2024)]%
        {Gartner5188263}
Gartner5188263 \bibinfo{year}{2024}\natexlab{}.
\newblock \bibinfo{title}{{How to Calculate Business Value and Cost for Generative AI Use Cases (\url{https://www.gartner.com/en/documents/5188263})}}.
\newblock
\newblock


\bibitem[Guo et~al\mbox{.}(2020)]%
        {Ruiqi20}
\bibfield{author}{\bibinfo{person}{Ruiqi Guo}, \bibinfo{person}{Philip Sun}, \bibinfo{person}{Erik Lindgren}, \bibinfo{person}{Quan Geng}, \bibinfo{person}{David Simcha}, \bibinfo{person}{Felix Chern}, {and} \bibinfo{person}{Sanjiv Kumar}.} \bibinfo{year}{2020}\natexlab{}.
\newblock \showarticletitle{{Accelerating Large-Scale Inference with Anisotropic Vector Quantization}}. In \bibinfo{booktitle}{\emph{Proceedings of the International Conference on Machine Learning (ICML)}}. \bibinfo{pages}{3887--3896}.
\newblock


\bibitem[J{\'{e}}gou et~al\mbox{.}(2011)]%
        {JegouDS11}
\bibfield{author}{\bibinfo{person}{Herv{\'{e}} J{\'{e}}gou}, \bibinfo{person}{Matthijs Douze}, {and} \bibinfo{person}{Cordelia Schmid}.} \bibinfo{year}{2011}\natexlab{}.
\newblock \showarticletitle{{Product Quantization for Nearest Neighbor Search}}.
\newblock \bibinfo{journal}{\emph{IEEE Transactions on Pattern Analysis and Machine Intelligence (TPAMI)}} \bibinfo{volume}{33}, \bibinfo{number}{1} (\bibinfo{year}{2011}), \bibinfo{pages}{117--128}.
\newblock


\bibitem[Malkov and Yashunin(2020)]%
        {HNSW}
\bibfield{author}{\bibinfo{person}{Yury~A. Malkov} {and} \bibinfo{person}{Dmitry~A. Yashunin}.} \bibinfo{year}{2020}\natexlab{}.
\newblock \showarticletitle{{Efficient and Robust Approximate Nearest Neighbor Search Using Hierarchical Navigable Small World Graphs}}.
\newblock \bibinfo{journal}{\emph{IEEE Transactions on Pattern Analysis and Machine Intelligence (TPAMI)}} \bibinfo{volume}{42}, \bibinfo{number}{4} (\bibinfo{year}{2020}), \bibinfo{pages}{824--836}.
\newblock


\bibitem[Mohoney et~al\mbox{.}(2023)]%
        {MohoneyPCMIMPR23}
\bibfield{author}{\bibinfo{person}{Jason Mohoney}, \bibinfo{person}{Anil Pacaci}, \bibinfo{person}{Shihabur~Rahman Chowdhury}, \bibinfo{person}{Ali Mousavi}, \bibinfo{person}{Ihab~F. Ilyas}, \bibinfo{person}{Umar~Farooq Minhas}, \bibinfo{person}{Jeffrey Pound}, {and} \bibinfo{person}{Theodoros Rekatsinas}.} \bibinfo{year}{2023}\natexlab{}.
\newblock \showarticletitle{{High-Throughput Vector Similarity Search in Knowledge Graphs}}.
\newblock \bibinfo{journal}{\emph{Proc. {ACM} Manag. Data}} \bibinfo{volume}{1}, \bibinfo{number}{2} (\bibinfo{year}{2023}), \bibinfo{pages}{197:1--197:25}.
\newblock


\bibitem[Neo4j({[n.\,d.]})]%
        {Neo4j}
Neo4j \bibinfo{year}{[n.\,d.]}\natexlab{}.
\newblock \bibinfo{title}{{Neo4j (\url{https://neo4j.com/})}}.
\newblock
\newblock


\bibitem[Neo4jVec({[n.\,d.]})]%
        {Neo4jVec}
Neo4jVec \bibinfo{year}{[n.\,d.]}\natexlab{}.
\newblock \bibinfo{title}{{Neo4j Vector Index and Search (\url{https://neo4j.com/labs/genai-ecosystem/vector-search/})}}.
\newblock
\newblock


\bibitem[Neptune({[n.\,d.]})]%
        {Neptune}
Neptune \bibinfo{year}{[n.\,d.]}\natexlab{}.
\newblock \bibinfo{title}{{Neptune (\url{https://aws.amazon.com/neptune/})}}.
\newblock
\newblock


\bibitem[NeptuneVec({[n.\,d.]})]%
        {NeptuneVec}
NeptuneVec \bibinfo{year}{[n.\,d.]}\natexlab{}.
\newblock \bibinfo{title}{{Vector Indexing in Neptune Analytics (\url{https://docs.aws.amazon.com/neptune-analytics/latest/userguide/vector-index.html})}}.
\newblock
\newblock


\bibitem[Pan et~al\mbox{.}(2024a)]%
        {VecDBSurvey24}
\bibfield{author}{\bibinfo{person}{James~Jie Pan}, \bibinfo{person}{Jianguo Wang}, {and} \bibinfo{person}{Guoliang Li}.} \bibinfo{year}{2024}\natexlab{a}.
\newblock \showarticletitle{{Survey of Vector Database Management Systems}}.
\newblock \bibinfo{journal}{\emph{{VLDB} Journal}} \bibinfo{volume}{33}, \bibinfo{number}{5} (\bibinfo{year}{2024}), \bibinfo{pages}{1591--1615}.
\newblock


\bibitem[Pan et~al\mbox{.}(2024b)]%
        {VecDBTutorial24}
\bibfield{author}{\bibinfo{person}{James~Jie Pan}, \bibinfo{person}{Jianguo Wang}, {and} \bibinfo{person}{Guoliang Li}.} \bibinfo{year}{2024}\natexlab{b}.
\newblock \showarticletitle{{Vector Database Management Techniques and Systems}}. In \bibinfo{booktitle}{\emph{Proceedings of the ACM International Conference on Management of Data (SIGMOD)}}. \bibinfo{pages}{597--604}.
\newblock


\bibitem[Peng et~al\mbox{.}(2024)]%
        {abs-2408-08921}
\bibfield{author}{\bibinfo{person}{Boci Peng}, \bibinfo{person}{Yun Zhu}, \bibinfo{person}{Yongchao Liu}, \bibinfo{person}{Xiaohe Bo}, \bibinfo{person}{Haizhou Shi}, \bibinfo{person}{Chuntao Hong}, \bibinfo{person}{Yan Zhang}, {and} \bibinfo{person}{Siliang Tang}.} \bibinfo{year}{2024}\natexlab{}.
\newblock \showarticletitle{{Graph Retrieval-Augmented Generation: {A} Survey}}.
\newblock \bibinfo{journal}{\emph{CoRR}}  \bibinfo{volume}{abs/2408.08921} (\bibinfo{year}{2024}).
\newblock


\bibitem[Pgvector({[n.\,d.]})]%
        {Pgvector}
Pgvector \bibinfo{year}{[n.\,d.]}\natexlab{}.
\newblock \bibinfo{title}{{pgvector (\url{https://github.com/pgvector/pgvector})}}.
\newblock
\newblock


\bibitem[Pinecone({[n.\,d.]})]%
        {Pinecone}
Pinecone \bibinfo{year}{[n.\,d.]}\natexlab{}.
\newblock \bibinfo{title}{{Pinecone (\url{https://www.pinecone.io/})}}.
\newblock
\newblock


\bibitem[Roberts(2024)]%
        {arizeRAG}
\bibfield{author}{\bibinfo{person}{Amber Roberts}.} \bibinfo{year}{2024}\natexlab{}.
\newblock \bibinfo{title}{{RAG Evaluation (\url{https://arize.com/blog-course/rag-evaluation/})}}.
\newblock
\newblock


\bibitem[Subramanya et~al\mbox{.}(2019)]%
        {DiskANN}
\bibfield{author}{\bibinfo{person}{Suhas~Jayaram Subramanya}, \bibinfo{person}{Fnu Devvrit}, \bibinfo{person}{Harsha~Vardhan Simhadri}, \bibinfo{person}{Ravishankar Krishnawamy}, {and} \bibinfo{person}{Rohan Kadekodi}.} \bibinfo{year}{2019}\natexlab{}.
\newblock \showarticletitle{{Rand-NSG: Fast Accurate Billion-point Nearest Neighbor Search on a Single Node}}. In \bibinfo{booktitle}{\emph{Annual Conference on Neural Information Processing Systems (NeurIPS)}}. \bibinfo{pages}{13748--13758}.
\newblock


\bibitem[Sz\'{a}rnyas et~al\mbox{.}(2022)]%
        {ldbc-snb}
\bibfield{author}{\bibinfo{person}{G\'{a}bor Sz\'{a}rnyas}, \bibinfo{person}{Jack Waudby}, \bibinfo{person}{Benjamin~A. Steer}, \bibinfo{person}{D\'{a}vid Szak\'{a}llas}, \bibinfo{person}{Altan Birler}, \bibinfo{person}{Mingxi Wu}, \bibinfo{person}{Yuchen Zhang}, {and} \bibinfo{person}{Peter Boncz}.} \bibinfo{year}{2022}\natexlab{}.
\newblock \showarticletitle{{The LDBC Social Network Benchmark: Business Intelligence Workload}}.
\newblock \bibinfo{journal}{\emph{Proceedings of the VLDB Endowment (PVLDB)}} \bibinfo{volume}{16}, \bibinfo{number}{4} (\bibinfo{year}{2022}), \bibinfo{pages}{877--890}.
\newblock


\bibitem[Tian(2022)]%
        {Tian22}
\bibfield{author}{\bibinfo{person}{Yuanyuan Tian}.} \bibinfo{year}{2022}\natexlab{}.
\newblock \showarticletitle{{The World of Graph Databases from An Industry Perspective}}.
\newblock \bibinfo{journal}{\emph{{SIGMOD} Record}} \bibinfo{volume}{51}, \bibinfo{number}{4} (\bibinfo{year}{2022}), \bibinfo{pages}{60--67}.
\newblock


\bibitem[Wang et~al\mbox{.}(2024)]%
        {VecDBPanel24}
\bibfield{author}{\bibinfo{person}{Jianguo Wang}, \bibinfo{person}{Shasank Chavan}, \bibinfo{person}{Guoliang Li}, \bibinfo{person}{Yannis Papakonstantinou}, {and} \bibinfo{person}{Charles Xie}.} \bibinfo{year}{2024}\natexlab{}.
\newblock \showarticletitle{{Vector Databases: What's Really New and What's Next?}}
\newblock \bibinfo{journal}{\emph{Proceedings of the VLDB Endowment (PVLDB)}} \bibinfo{volume}{17}, \bibinfo{number}{12} (\bibinfo{year}{2024}), \bibinfo{pages}{4505--4506}.
\newblock


\bibitem[Wang et~al\mbox{.}(2021)]%
        {Milvus21}
\bibfield{author}{\bibinfo{person}{Jianguo Wang}, \bibinfo{person}{Xiaomeng Yi}, \bibinfo{person}{Rentong Guo}, \bibinfo{person}{Hai Jin}, \bibinfo{person}{Peng Xu}, \bibinfo{person}{Shengjun Li}, \bibinfo{person}{Xiangyu Wang}, \bibinfo{person}{Xiangzhou Guo}, \bibinfo{person}{Chengming Li}, \bibinfo{person}{Xiaohai Xu}, \bibinfo{person}{Kun Yu}, \bibinfo{person}{Yuxing Yuan}, \bibinfo{person}{Yinghao Zou}, \bibinfo{person}{Jiquan Long}, \bibinfo{person}{Yudong Cai}, \bibinfo{person}{Zhenxiang Li}, \bibinfo{person}{Zhifeng Zhang}, \bibinfo{person}{Yihua Mo}, \bibinfo{person}{Jun Gu}, \bibinfo{person}{Ruiyi Jiang}, \bibinfo{person}{Yi Wei}, {and} \bibinfo{person}{Charles Xie}.} \bibinfo{year}{2021}\natexlab{}.
\newblock \showarticletitle{{Milvus: {A} Purpose-Built Vector Data Management System}}. In \bibinfo{booktitle}{\emph{Proceedings of the ACM International Conference on Management of Data (SIGMOD)}}. \bibinfo{pages}{2614--2627}.
\newblock


\bibitem[Wei et~al\mbox{.}(2020)]%
        {analyticdb}
\bibfield{author}{\bibinfo{person}{Chuangxian Wei}, \bibinfo{person}{Bin Wu}, \bibinfo{person}{Sheng Wang}, \bibinfo{person}{Renjie Lou}, \bibinfo{person}{Chaoqun Zhan}, \bibinfo{person}{Feifei Li}, {and} \bibinfo{person}{Yuanzhe Cai}.} \bibinfo{year}{2020}\natexlab{}.
\newblock \showarticletitle{{AnalyticDB-V: {A} Hybrid Analytical Engine Towards Query Fusion for Structured and Unstructured Data}}.
\newblock \bibinfo{journal}{\emph{Proceedings of the VLDB Endowment (PVLDB)}} \bibinfo{volume}{13}, \bibinfo{number}{12} (\bibinfo{year}{2020}), \bibinfo{pages}{3152--3165}.
\newblock


\bibitem[Yang et~al\mbox{.}(2020)]%
        {PASE20}
\bibfield{author}{\bibinfo{person}{Wen Yang}, \bibinfo{person}{Tao Li}, \bibinfo{person}{Gai Fang}, {and} \bibinfo{person}{Hong Wei}.} \bibinfo{year}{2020}\natexlab{}.
\newblock \showarticletitle{{PASE: PostgreSQL Ultra-High-Dimensional Approximate Nearest Neighbor Search Extension}}. In \bibinfo{booktitle}{\emph{Proceedings of the ACM International Conference on Management of Data (SIGMOD)}}. \bibinfo{pages}{2241--2253}.
\newblock


\bibitem[Zhang et~al\mbox{.}(2024)]%
        {VecDBRDBMS}
\bibfield{author}{\bibinfo{person}{Yunan Zhang}, \bibinfo{person}{Shige Liu}, {and} \bibinfo{person}{Jianguo Wang}.} \bibinfo{year}{2024}\natexlab{}.
\newblock \showarticletitle{{Are There Fundamental Limitations in Supporting Vector Data Management in Relational Databases? A Case Study of PostgreSQL}}. In \bibinfo{booktitle}{\emph{Proceedings of the International Conference on Data Engineering (ICDE)}}. \bibinfo{pages}{2835--2849}.
\newblock


\end{thebibliography}

\end{document}